\documentclass{optica-article}
\journal{opticajournal} 

\articletype{Research Article}
\usepackage{xspace}
\usepackage{amsmath}
\usepackage{graphicx}
\usepackage{color}
\definecolor{darkred}{rgb}{0.8, 0.0, 0.0}
\definecolor{darkblue}{rgb}{0.2, 0.2, 0.6}
\newcommand{\etal}{\textit{et al.\@}\xspace}
\newcommand{\exvivo}{\textit{ex vivo}\xspace}

\newcommand{\invivo}{\textit{in vivo}\xspace}
\newcommand{\Invivo}{\textit{In vivo}\xspace}
\newcommand{\invitro}{\textit{in vitro}\xspace}

\newcommand{\enface}{\textit{en face}\xspace}
\newcommand{\um}{\(\muup\)m\xspace}
\newcommand{\uM}{\(\muup\)M\xspace}

\newcommand{\cbum}{\(\muup\)m$^3$\xspace}
\newcommand{\sqdb}{dB$^2$\xspace}
\newcommand{\invs}{s$^{-1}$\xspace}

\newcommand{\LIV}{\mathrm{LIV}\xspace}
\newcommand{\OCDSe}{OCDS$_e$\xspace}
\newcommand{\OCDSl}{OCDS$_l$\xspace}
\newcommand{\aLIV}{\mathrm{aLIV}\xspace}

\newcommand{\Swiftness}{\mathrm{Swiftness}\xspace}

\newcommand{\DSR}{DSR\xspace}
\newcommand{\DSRs}{DSRs\xspace}

\newcommand{\VAtw}{{V\!\! A_{tw}\xspace}}
\newcommand{\Atw}{{A_{tw}\xspace}}
\newcommand{\tw}{{T_w\xspace}}

\newcommand{\revstart}[1]{{}}
\newcommand{\revend}{}
	
\graphicspath{{./figures}{.}{./figures_tmp}}

\begin{document}

\title{Dynamic optical coherence tomography algorithm for label-free assessment of swiftness and occupancy of intratissue moving scatterers}

\author{Rion Morishita\authormark{1},
		Pradipta Mukherjee\authormark{1,2},
		Ibrahim Abd El-Sadek\authormark{1,3},
		Tanatchaya Seesan\authormark{1,4},
		Tomoko Mori\authormark{5},
		Atsuko Furukawa\authormark{5},
		Shinichi Fukuda\authormark{6,7},
		Donny Lukmanto\authormark{6},
		Satoshi Matsusaka\authormark{5},
		Shuichi Makita\authormark{1},
		and Yoshiaki Yasuno\authormark{1,*}
		}

\address{\authormark{1}Computational Optics Group, University of Tsukuba, Tsukuba, Ibaraki, Japan.\\
		 \authormark{2}Centre for Biomedical Engineering, Indian Institute of Technology Delhi, New Delhi, India.\\
		 \authormark{3}Department of Physics, Faculty of Science, Damietta University, New Damietta City, Damietta, Egypt.\\
		 \authormark{4}School of Science, King Mongkut’s Institute of Technology Ladkrabang, Bangkok, Thailand.\\
	 	 \authormark{5}Clinical Research and Regional Innovation, Faculty of Medicine, University of Tsukuba, Tsukuba, Ibaraki, Japan.\\
 	 	 \authormark{6}Laboratory of Advanced Vision Science, Faculty of Medicine, University of Tsukuba, Tsukuba, Ibaraki, Japan.\\
 	 	 \authormark{7}Department of Ophthalmology, Faculty of Medicine, University of Tsukuba, Ibaraki, Japan.\\
  }

\email{\authormark{*}yoshiaki.yasuno@cog-labs.org}

\begin{abstract*}
Dynamic optical coherence tomography (DOCT) statistically analyzes fluctuations in time-sequential OCT signals, enabling label-free and three-dimensional visualization of intratissue and intracellular activities.
Current DOCT methods, such as logarithmic intensity variance (LIV) and OCT correlation decay speed (OCDS) have several limitations.
Namely, the DOCT values and intratissue motions are not directly related, and hence DOCT values are not interpretable in the context of the tissue motility.
We introduce an open-source DOCT algorithm that provides more direct interpretation of DOCT in the contexts of dynamic scatterer ratio and scatterer speed in the tissue.
The detailed properties of the new and conventional DOCT methods are investigated by numerical simulations based on our open-source DOCT simulation framework, and the experimental validation with \invitro and \exvivo samples demonstrates the feasibility of the method.
\end{abstract*}

\section{Introduction}
Optical coherence tomography (OCT) \cite{Huang1991Science} is a low-coherence interferometric imaging modality, and the advantages of OCT over conventional microscopic techniques have recently led to its use in microscopic investigations \cite{Izatt1994OL, Beaurepaire1998OL, Drexler1999OL, Dubois2004AO, Aguirre2015Text, Huang2019JVE, SELin2021BOE, Ming2022Biosens, KYChen2023BOE, KYChen2024SciRep}.
First, OCT uses a near-infrared light probe, which provides cellular-level imaging at deeper regions than conventional microscopy, such as at around 1-mm depth.
Second, OCT employs the endogenous scattering of the sample as the source of contrast.
Hence, OCT is label-free and non-invasive.

One limitation of OCT is its lack of sensitivity to intratissue and intracellular activities.
However, this limitation has recently been overcome by an emerging methodology called dynamic OCT (DOCT) \cite{Apelian2016BOE, Ren2024CB}.
DOCT is a combination of time-sequential OCT signal acquisition and subsequent temporal analysis of the signal sequence.

Several signal analysis algorithms have been developed for DOCT.
One type of algorithm computes the variance or standard deviation of the time-sequential OCT signals, directly \cite{Oldenburg2012BOE, Oldenburg2015Optica, Apelian2016BOE, Thouvenin2017JBO, El-Sadek2020BOE, Park2021BOE} or indirectly \cite{Scholler2019OpEx, KYFei2024BOE}.
This approach uses the magnitude of the OCT signal fluctuations to contrast intratissue and intracellular activities. 
Another type of algorithm computes the temporal autocorrelation of time-sequential OCT signals \cite{Oldenburg2015Optica, Apelian2016BOE,El-Sadek2020BOE}, and uses the decorrelation properties as the contrast source.
\revstart{c1-3}
Yet another type of algorithm analyzes the power spectral density of time-sequential OCT signals through a numerical Fourier transform \cite{Ling2017LSM, Thouvenin2017IOVS, Munter2020OL, Leung2020BOE, Xia2023Optica, KYChen2024BOE, Yin2025NPJ}, and uses the dominant frequency or powers within several frequency bins as the contrast source.
The final type of algorithm analyzes the phase difference between the complex OCT signals at two adjacent time points \cite{Hildebrandt2025OL}, and uses the average and standard deviation of all the phase differences as the contrast source.
\revend

The present authors have previously developed two DOCT algorithms.
The first is the logarithmic intensity variance (LIV), which is defined as the time variance of dB-scaled OCT signal intensity \cite{El-Sadek2020BOE}.
LIV quantifies the magnitude of signal scintillation, which had been considered to be related to the magnitude of intratissue activities.
The other is the OCT correlation decay speed (OCDS), which is defined as the slope of the auto-correlation curve of the time-sequential OCT signal within a particular range of the auto-correlation delay time \cite{El-Sadek2020BOE}.
OCDS reflects the speed of OCT signal scintillation, which had been considered to be related to the speed of the intratissue activities.
Previous studies have examined delay-time ranges of [12.8 ms, 64 ms] and [204.8 ms, 1228.8 ms], with the corresponding OCDS referred to as early OCDS (\OCDSe) and late OCDS (\OCDSl), respectively \cite{El-Sadek2020BOE, El-Sadek2021BOE}.
\OCDSe and \OCDSl are considered to be indicators of fast and slow intratissue activities, respectively.

LIV and OCDS have been applied to various biological samples.
For tumor spheroids, LIV and OCDS can reveal the alterations in intratissue activities induced by anti-cancer drugs \cite{El-Sadek2021BOE, El-Sadek2023SciRep,  El-Sadek2024SciRep}.
For alveolar organoids, LIV has revealed heterogeneous cellular activities within the alveolar epithelium, which may indicate abnormal reprogramming of the epithelial cells, a phenomenon known as bronchiolization \cite{Morishita2023BOE}. 
For \exvivo mouse organs, LIV has revealed metabolic activities in livers \cite{Mukherjee2021SciRep, Mukherjee2022BOE} and kidneys \cite{Mukherjee2023SciRep}.
LIV has also been successfully applied to \invivo human skin, \invitro skin models, and \exvivo skin samples \cite{Guo2025arXiv}.

Despite their feasibility and utility, the LIV and OCDS algorithms have two shortcomings.
First, OCDS is not a direct indicator of the speed of dynamics. 
The selection of the delay-time range is arbitrary, and the correlation decay slope within a specific delay range neither directly nor monotonically correlates with the speed of the dynamics.
For example, in drug testing, the responses of \invitro samples to various drug types and doses are evaluated.
This requires a modality capable of sensitively distinguishing between different types of cellular activities.
We believe that biological conditions can be effectively differentiated by identifying the speed of intratissue dynamics.
\revstart{c1-1}
Similar to OCDS, frequency-based DOCT methods generate contrast values relating to the fluctuation speed of OCT signals.
However, the frequency of OCT signal is not a direct measure of the motion speed of the intratissue dynamic.
\revend
Second, these DOCT methods quantify the dynamics of OCT signals rather than the dynamics of tissues.
The relationship between the OCT-signal dynamics and tissue dynamics remains unclear.
These two shortcomings hinder the interpretation of LIV and OCDS images in the context of tissue activities.

This paper addresses these two issues.
The first issue is solved by introducing two new DOCT metrics (i.e., contrasts), which are computed simultaneously from an OCT time sequence.
The first metric, ``Swiftness'', is a more direct indicator of the speed of tissue dynamics than OCDS.
The second metric, ``authentic LIV (aLIV)'', is a byproduct of Swiftness, and provides an alternative to LIV.
The second issue is solved by numerically analyzing the relationship between tissue dynamics and various DOCT metrics, including LIV, OCDS, Swiftness, and aLIV.
For the numerical simulations, we conduct a mathematical model of dynamic tissues as an extension of our previous dispersed scatterer model (DSM) \cite{Tomita2023BOE}.
Briefly, this paper consists of three main components: the principles of the new DOCT metrics (Section \ref{sec:principle}), a numerical study investigating their properties (Section \ref{sec:numerical simulation}), and an experimental study validating their applicability to biological samples (Section \ref{sec:experimental validation}).
For the experimental study, \invitro tumor spheroids and an \exvivo mouse kidney are used as samples.

\section{New DOCT algorithm}
\label{sec:principle}
Our new open-source DOCT algorithm \cite{VlivGitHub} uses the dependency of conventional LIV on the size of the acquisition-time-window ($\Atw$) \cite{El-Sadek2020BOE}.
LIV is defined as the time variance of the time-sequential OCT signal.
Thus, LIV becomes larger or smaller in proportion to the fluctuation magnitude of the OCT signal.
However, LIV is also influenced by the size of $\Atw$; specifically, LIV (i.e., the time variance) inevitably decreases as $\Atw$ becomes smaller.

\begin{figure}
	\centering\includegraphics{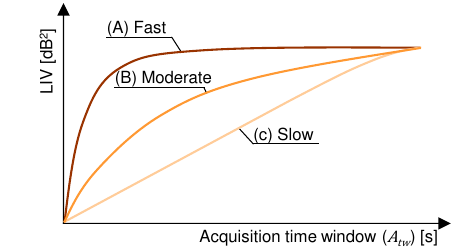}
	\caption{
		Schematic illustrating the acquisition-time-window ($\Atw$)-size dependency of LIV.
		LIV inevitably decreases as $\Atw$ becomes smaller.
		Curves A, B, and C represent the cases of faster, moderate, and slower signal fluctuations, respectively.}
	\label{fig:schematic LIV curve}
\end{figure}
Figure \ref{fig:schematic LIV curve} schematically illustrates the properties of LIV, where curves A, B, and C represent the cases of faster, moderate, and slower signal fluctuations, respectively.
LIV increases as $\Atw$ becomes larger, and then saturates when $\Atw$ is sufficiently large in respect to the signal fluctuation cycle.

In this paper, we propose new DOCT metrics (contrasts) by leveraging this property of LIV.
One metric, related to the speed of intratissue activity, is defined from the salutation speed of the LIV-$\Atw$ curve.
The other metric, quantifying the magnitude of OCT signal fluctuations, is defined by the saturation level of the curve.
The theory and algorithm are described in detail in the following sections. 

\subsection{LIV curve}
\label{subsec: LIV curve}
\begin{figure}
	\centering\includegraphics{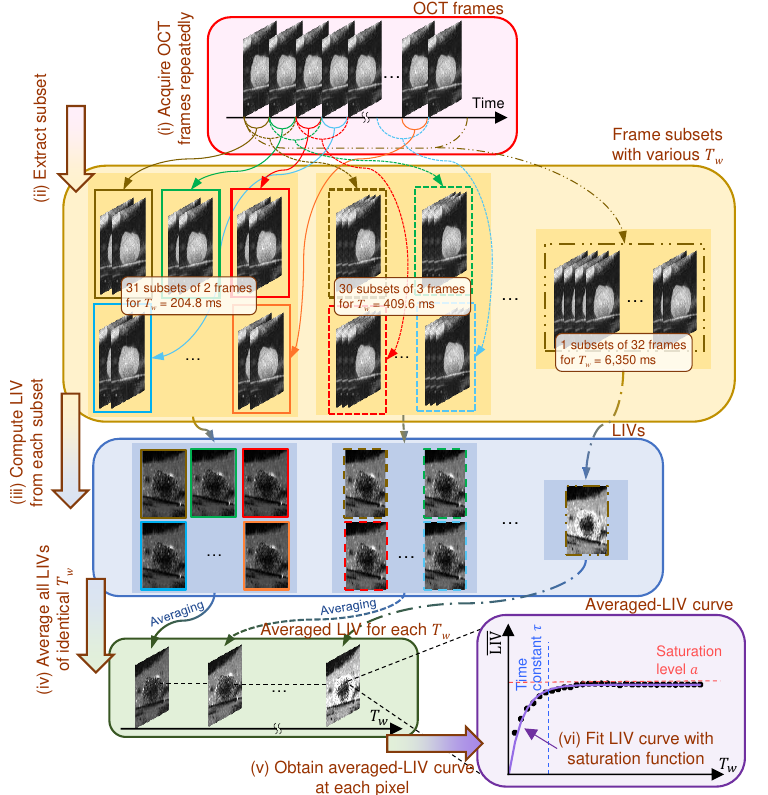}
	\caption{
		Diagram showing the principle of the new DOCT algorithm.
		(i) OCT frames are repeatedly acquired with a time interval of $\Delta t$ over $\Atw$.
		(ii) All combinations of data subsets are extracted for different time window ($\tw$) sizes from the time-sequential OCT frames.
		(iii) LIV is computed from each data subset.
		(iv) All LIV values with identical $\tw$ are averaged.
		(v) A curve of the averaged LIV over $\tw$ is obtained, which is referred to as the ``LIV curve''.
		(vi) The LIV curve is fitted with a first-order saturation function of $\tw$.
	}
	\label{fig:schematic principle}
\end{figure}
In the new DOCT method, similar to conventional DOCT imaging, OCT frames are repeatedly acquired with a time interval of $\Delta t$ over $\Atw$ at the same sample location, as schematically illustrated in (i, red box) of Fig.\@ \ref{fig:schematic principle}.

Several LIV values with different time window are then computed to form the LIV-$\tw$ curve.
For this computation, at first, data subsets of different time window ($\tw$) sizes are extracted from the time-sequential OCT signal [(ii) of Fig.\@ \ref{fig:schematic principle}].
For clarification, the total acquisition time duration for the time-sequential OCT signal is the acquisition time window ($\Atw$), while the time duration of each data subset used to compute an LIV value is the time window ($\tw$).

Subsequently, as illustrated in (iii) of Fig.\@ \ref{fig:schematic principle}, the LIV values of each data subset are calculated in the same way as for conventional LIV \cite{El-Sadek2020BOE, El-Sadek2021BOE}, i.e.,
\begin{equation}
	\label{eq:LIV}
	\LIV(x,z) = \frac{1}{N}\sum_{i=0}^{N-1}\left[\mathrm{I}_\mathrm{dB}(x, z, t_i)-\langle \mathrm{I}_\mathrm{dB}(x, z, t_i) \rangle_t\right]^2,
\end{equation}
where $\mathrm{I}_\mathrm{dB}(x, z, t_i)$ represents the dB-scaled OCT signal intensity at a depth of $z$ and a lateral position of $x$ in the OCT frame.
$t_i$ is the sampling time of the $i$-th OCT frame, where $i = 0, 1, 2, \cdots, N-1$, and $N$ is the total number of frames in the data subset.
$\langle\quad\rangle_t$ denotes averaging over time.
The smallest $\tw$ corresponds to the inter-frame interval $\Delta t = t_i - t_{i-1}$ of the time-sequential OCT frames, while the largest $\tw$ is equal to $\Atw$.

Note that two or more data subsets with identical $\tw$ can be obtained, except for the largest $\tw$.
In the cases where multiple data subsets share the same $\tw$, the LIV values for all subsets are computed and then averaged [(iv) of Fig.\@ \ref{fig:schematic principle}] as
\begin{equation}
	\label{eq:average LIV}
	\overline{\LIV}(x, z; \tw) = \frac{1}{N_{\tw}}\sum_{l=0}^{N_{\tw}-1}\LIV_l(x, z; \tw),
\end{equation}
where $\overline{\LIV}(x, z; \tw)$ is the averaged LIV for a specific $\tw$ and $\LIV_l(x, z; \tw)$ represents the $l$-th LIV among all LIV values with identical $\tw$, where $l = 0, 1, 2, \cdots, N_{\tw}-1$.
$N_{Tw}$ represents the total number of LIV values corresponding to identical $\tw$.
This averaged LIV is used as the representative LIV of that particular time window.

The averaged LIVs are computed for all possible $\tw$ values and, at each pixel, a curve of the averaged LIV is obtained as a function of $\tw$, as schematically depicted in (v) of Fig.\@ \ref{fig:schematic principle}.
This curve of the averaged LIV over $\tw$ is hereafter referred to as the ``LIV curve''.

\subsection{Authentic LIV and Swiftness}
To extract the characteristic parameters, the LIV curve obtained using the method described in Section \ref{subsec: LIV curve}, i.e., the curve of the averaged LIV along $\tw$ at each pixel, is fitted with a first-order saturation function of $\tw$ [(vi) in Fig.\@ \ref{fig:schematic principle}]:
\begin{equation}
	\label{eq:1st-order saturation function}
	f_{\LIV}(\tw;a,\tau) = a\left[1-\exp\left(-\tw/\tau\right)\right],
\end{equation}
where the fitting parameters $a$ and $\tau$ represent the saturation level and the time constant, respectively.

We define two DOCT metrics using these fitting parameters.
The first metric is the ``authentic LIV (aLIV),'' which is defined as
\begin{equation}
	\label{eq:aLIV}
	\aLIV \equiv a.
\end{equation}
In principle, aLIV is expected to be an $\Atw$-independent version of LIV.

The second metric, ``Swiftness,'' is defined as 
\begin{equation}
	\label{eq:swiftness}
	\Swiftness \equiv \frac{1}{\tau}.
\end{equation}
Swiftness quantifies the saturation speed of the LIV curve.

It should be noted that both aLIV and Swiftness are direct measures of the fluctuation of the OCT signal, rather than the intratissue dynamics.
The relationships between the intratissue dynamics and these DOCT metrics will be discussed in Section \ref{sec:numerical simulation}.

\subsection{Implementation of aLIV and Swiftness}
\label{subsec:implementation new DOCT}
In our implementation, the new DOCT algorithm is applied to OCT frames that are acquired 32 times with an inter-frame time of 204.8 ms (see Section \ref{subsubsec:OCT and DOCT} for details).
Namely, the smallest $\tw$ is 204.8 ms (corresponding to the two-frame subsets) and the largest $\tw$ is 6.3 s (corresponding to the 32-frame subset).

To improve the accuracy of fitting the LIV curve, the LIV curve is smoothed by applying 3 $\times$ 3 kernel averaging to each averaged LIV cross-section $\overline{\LIV}(x, z; \tw)$ at each $\tw$.
The kernel averaging is implemented using the OpenCV function ``cv2.blur().''

For the function fitting, the Levenberg-Marquardt algorithm is used to solve the nonlinear least-squares problem.
The first-order saturation function in Eq.\@ (\ref{eq:1st-order saturation function}) is used as the model function and the optimal values of $a$ and $\tau$ within the range $\left[0,\infty\right]$ are determined.

The processing for the new DOCT algorithm was executed on a laptop PC equipped with an Intel Core i7-10750H CPU and 32 GB memory, running the Windows 10 Pro operating system.
The PC was also equipped with a graphic processing unit (GPU), specifically NVIDIA GeForce RTX 2070 Super with Max-Q Design.

The new DOCT algorithm was implemented in Python 3.8.5 using NumPy 1.19.2 and CuPy 10.0.0 libraries for all computations except function fitting.
To accelerate the function fitting, the GPU and an open-source GPU-accelerated implementation of function fitting (Gpufit \cite{Przybylski2017SciRep}) were used.
The LIV curve was fitted using the ``gf.fit\_constrained()'' function in the Gpufit 1.2.0 library.
The first-order saturation function [Eq.\@ (\ref{eq:1st-order saturation function})] is not included in the built-in model functions of Gpufit library.
Thus, we implemented the function module and recompiled the library with this module.
The Python implementation of the new DOCT algorithm is available as open source at the GitHub repository \cite{VlivGitHub}.
Details regarding the processing time are discussed in Section \ref{subsec:processing time}.

\section{Numerical investigation of DOCT properties}
\label{sec:numerical simulation}
The relationships between the intratissue dynamics and DOCT metrics were investigated through numerical simulations using our open-source DOCT simulation framework \cite{DoctSimuGitHub}.

\subsection{Simulation theory and method}
\subsubsection{Dynamic sample modeling}
\label{sec:sampleModel}
We modeled a sample based on dispersed scatterer model (DSM) \cite{Tomita2023BOE}, which represents a sample as a spatially slowly varying refractive index distribution with infinitely small scatterers randomly dispersed throughout the sample.
In our simulations, the slowly varying refractive indices were simplified to be constant, and the refractive indices of all the scatterers were assumed to be identical.
Hence, the all scatterers had the same intensity reflectivity.
It should be noted that the sample represented by this model does not have macroscopic inhomogeneity (i.e., large structure).
However, each DOCT value is computed from a small region, as small as the size of the point spread function (PSF) of OCT.
And hence, this lack of large structure is not a significant limitation of the simulation.

The intratissue scatterers were assumed to follow not Brownian motion (i.e., diffusion) but random ballistic motion in which all scatterers move rectilinearly with the same constant speed $v$, but with a motion direction that is randomly chosen.
This selection of random ballistic motion was based on an extensive literature survey about the intratissue and intracellular activities and simulation results using the motion models \cite{Feng2025BOE}.
Further points about the motion model will be discussed in Section \ref{sec:discuss_complicateModel}.
Assuming the initial position of the $j$-th scatterer is ($x_{j0}, y_{j0}$, $z_{j0}$), the three-dimensional (3D) position of the scatterer after a particular travel time $t$ is given by
\begin{equation}
	\label{eq:scatterer location}
	\begin{split}
		&x_j(t;\varphi_j,\theta_j) = x_{j0} + vt\sin\varphi_j\cos\theta_j,\\
		&y_j(t;\varphi_j,\theta_j) = y_{j0} + vt\sin\varphi_j\sin\theta_j,\\
		&z_j(t;\varphi_j,\theta_j) = z_{j0} + vt\cos\varphi_j,\\
	\end{split}
\end{equation}
where $\varphi_j$ and $\theta_j$ are the azimuth and inclination angles, respectively, of the $j$-th scatterer's motion.
The position of the $j$-th scatterer at the $i$-th time point is expressed in incremental form as
\begin{equation}
	\label{eq:increment scattererLoc}
	\begin{split}
		&x_j(t_{i};\varphi_j,\theta_j) = x_j(t_{i-1};\varphi_j,\theta_j) + v\Delta t\sin\varphi_j\cos\theta_j,\\
		&y_j(t_{i};\varphi_j,\theta_j) = y_j(t_{i-1};\varphi_j,\theta_j) + v\Delta t\sin\varphi_j\sin\theta_j,\\
		&z_j(t_{i};\varphi_j,\theta_j) = z_j(t_{i-1};\varphi_j,\theta_j) + v\Delta t\cos\varphi_j,\\
	\end{split}
\end{equation}
where $\Delta t = t_i - t_{i-1}$ is the time interval between successive time points.

In the analyses in Sections \ref{sec:numericalProtocol} and \ref{sec:numericalResults}, we investigate the situation where dynamic scatterers and static scatterers coexist.
In this case, a portion of the scatterers move according to Eq.\@ (\ref{eq:scatterer location}) (or equivalently according to Eq.\@ (\ref{eq:increment scattererLoc})), while the other scatterers do not move.

\subsubsection{DOCT signal modeling}
\label{sec:signalModel}
At a single point in an image, the complex OCT signal obtained from the modeled sample is expressed as a summation of electric field contributions from each scatterer as
\revstart{c2-6}
\begin{equation}
	\label{eq:single complex OCT}
	\begin{split}
	E(t) \propto &\sum_{j=0}^{N_s-1} \color{darkblue}{b_j \exp\left[i\frac{2\pi}{\lambda}2z_j(t) + i\frac{2\pi}{\lambda}2\Delta z \right]} \\
	&\color{darkred}{\exp\left[-\frac{1}{2}\left(\frac{x_j(t)}{\sigma_x}\right)^2-\frac{1}{2}\left(\frac{y_j(t)}{\sigma_y}\right)^2-\frac{1}{2}\left(\frac{z_j(t)}{\sigma_z}\right)^2\right]},\\
\end{split}
\end{equation}
\revend
where $j$ is the scatterer index and $N_s$ is the number of scatterers in the sample.
The term $b_j$ and the first exponential part (blue part) correspond to the complex reflectivity of the $j$-th scatterer.
Here, $b_j$ is a coefficient accounting for the different reflectivity values of the scatterers, and we assume that this is 1 for all scatterers.
The exponential part represents the phase defined by the depth position of the scatterer, where $\lambda$ is the center wavelength of the probe beam and $\Delta z$ is the path length offset, defined by the arm lengths of the interference.
$\Delta z$ is a constant for all scatterer contributions, and therefore it does not affect the intensity of the OCT signal.

The second exponential (red) represents the 3D Gaussian PSF centered at ($x, y, z$) = (0, 0, 0), and $(x_j, y_j, z_j)$ is the position of the $j$-th scatterer.
Namely, the contributions from each scatterer are weighted by the PSF.
Let $\sigma_x, \sigma_y$, and $\sigma_z$ denote the width of the amplitude PSF defined by the standard deviation of the Gaussian.
The standard deviation and the $1/e^2$ width and full width at half maximum (FWHM) of the intensity of the OCT signal are related according to
\begin{equation} 
	\label{eq:sigma}
	\begin{split}
		&\sigma = \frac{1}{2\sqrt{2}}w_\mathrm{e},\\
		&\sigma = \frac{1}{2\sqrt{\ln2}}w_\mathrm{FWHM},\\
	\end{split}
\end{equation}
where $\sigma$ is the standard deviation width of the amplitude PSF and $w_\mathrm{e}$, $w_\mathrm{FWHM}$ are the $1/e^2$ width and FWHM of the OCT signal intensity, respectively.

Finally, the intensity of the OCT signal is obtained as
\begin{equation}
	\label{eq:intensity}
	I(t) = E(t)E^{*}(t),
\end{equation}
where $^{*}$ denotes the complex conjugate.
 
\subsubsection{Numerical generation of OCT time sequence}
The time sequence of the OCT intensity signal at a single point in the image is numerically simulated based on the theories described in Sections \ref{sec:sampleModel} and \ref{sec:signalModel}.

Specifically, we first generated a 3D numerical analysis field and randomly seeded the scatterers in the field.
As stated in Section \ref{sec:sampleModel}, we assumed that all scatterers had the same speed $|v|$, but random motion directions.
We computed the complex OCT signal at a single position in the image from the numerical field using Eq.\@ (\ref{eq:single complex OCT}), and then computed the OCT intensity from Eq.\@ (\ref{eq:intensity}).
After computing the OCT intensity at a time point, the positions of the scatterers were updated based on Eq.\@ (\ref{eq:increment scattererLoc}), and the OCT intensity at the new time point was computed.
By repeating this process, the time sequence of OCT intensity at a single position in an image was obtained. 

A scatterer that enters the analytic field during the simulation time, but had an initial position outside the analytic field, would not have been numerically seeded, and hence, would not contribute to the simulation results.
This reduces the simulation accuracy.
To lessen the effect of this inaccuracy, the length of the 3D analysis field is set to be $5\sigma_{\{x,y,z\}} + 2|v|\Atw$ for each direction, where the subscript $\{x,y,z\}$ corresponds to one of $x$, $y$, or $z$.
Namely, the field extends for $5 \sigma$-width of the complex PSF plus the maximum travel distance of a scatterer at both sides (see Supplementary Fig.\@ S1 for schematic depiction).

The OCT system parameters used for the simulations are identical to those of the OCT device used in the experimental validation described in Section \ref{sec:experimental validation}, i.e., $\lambda$ = 1.3 \um, the lateral resolution is 18 \um ($1/e^2$-width of intensity), and the axial resolution is 14 \um (FWHM).
The time-sequential OCT intensity was generated with a time interval (i.e., OCT frame interval) of 204.8 ms for 32 time points, therefore, $\Atw$ = 6.35 s, i.e., 204.8 ms $\times$ ($32-1$).
These parameters are consistent with the experiment described in Section \ref{subsubsec:OCT and DOCT}.

\subsubsection{Computation of DOCT values}
\label{subsubsec: simulation_DOCT method}
Four DOCT metrics, including the proposed aLIV and Swiftness metrics and our conventional LIV and OCDS metrics, were computed from the simulated time sequence of OCT intensity.

The proposed aLIV and Swiftness metrics were computed following the method described in Section \ref{sec:principle}.
In the simulations, nine LIV curves were averaged before performing the LIV curve fitting to emulate the $3\times3$-spatial-kernel averaging, which is one of the processing procedures explained in Section \ref{subsec:implementation new DOCT}.

The conventional metrics of LIV and OCDS were computed using the method described in Ref.\@ \cite{El-Sadek2020BOE, El-Sadek2021BOE}.
Namely, the time variance of the dB-scaled OCT intensity [Eq.\@ (\ref{eq:LIV})] was computed as the LIV, while the slope of the auto-correlation function of the dB-scaled OCT intensity at a delay-time range of [204.8 ms, 1228.8 ms] was computed as OCDS.
This delay-time range is the same as that of \OCDSl in Ref.\@ \cite{El-Sadek2021BOE}, and \OCDSl is known to be sensitive to relatively slow activity.
In the simulations, LIV and OCDS were obtained by averaging nine LIV and OCDS values to maintain consistency with the $3\times3$-kernel averaging in the aLIV and Swiftness computations.

It should be noted that, in real measurements (i.e., experiments), DOCT signals can be affected by fluctuations in the superior tissues, similar to the well-known projection artifact in OCT angiography \cite{Spaide2015Retina}.
This effect is not accounted for in our simulation.
Since the purpose of the simulation is to clarify the relationship between the DOCT signals and the local motion of the sample, this omission is not a significant limitation.

\subsection{Numerical study protocol}
\label{sec:numericalProtocol}
We conducted two numerical studies.
One investigated the dependencies of the DOCT metrics on the scatterer speed, while the other examined the dependency of the DOCT metrics on the proportion of dynamic scatterers.
The study protocols are described in the following subsections.

In addition to the newly introduced metrics (i.e., aLIV and Swiftness), previously established metrics, including LIV and OCDS, were also incorporated in each study.
This is because several \invitro and \exvivo samples have been analyzed using LIV and OCDS, with many carefully interpreted through comparisons with standard histology and fluorescence micrographs.
Comparing the new and existing DOCT metrics may facilitate the interpretation of images obtained using the newly introduced metrics.

\subsubsection{Study 1: Scatterer-speed dependency}
The first study investigated the scatterer-speed dependency of the DOCT metrics.
The simulations were conducted for 200 different scatterer speeds, selected from a range of [1.0 nm/s, 6.0 \um/s].
These 200 different speeds were not equally spaced, but were selected to clearly highlight the characteristics of the speed dependency. 
The minimum speed of the range (1.0 nm/s) was selected to cover the practical speed range of intratissue/intracellular motions which is from around 10 nm/s to 10 \um/s \cite{Feng2025BOE, Nolte2024RPP}.
Although the maximum speed of the simulation (6.0 \um/s) is slower than the maximum speed in the practical tissues/cells (10 \um/s), it results in a total displacement of 37.8 \um over the  simulation time window of 6.3 s, which is slightly larger than twice the lateral resolution of 18 \um.
And hence, the selected speed range is reasonable.
The speed was the same for all scatterers, but the direction of each scatterer was randomly selected.

The scatterer density was set to 0.055 scatterers/\cbum based on previously measured scatterer densities of tumor spheroids using a neural-network-based scatterer density estimator (NN-SDE) \cite{Seesan2021BOE, Seesan2024BOE}.
The NN-SDE is a NN model that processes a local OCT speckle pattern and estimates the scatterer density of the sample.
Both in the present numerical simulation and in the principle of the NN-SDE, the scatterer density is interpreted as the ``density of effective scatterers,'' where the effective scatterer is the scatterer whose scattering contributes to the OCT signal.

From the simulated time sequence of OCT intensity at each speed, the four DOCT metrics of aLIV, Swiftness, LIV, and OCDS were computed.
For each speed, five simulation trials were performed, and the median of each DOCT metric was used as the result.

\subsubsection{Study 2: Dynamic-scatterer ratio dependency}
In the second study, we assumed that a portion of the scatterers exhibit random ballistic motion, while the other scatterers remain static, i.e., not moving.
The dependency of the DOCT metrics on the dynamic scatterer ratio was investigated.

The dynamic-scatterer ratio (\DSR) is defined as
\begin{equation}
	\label{eq:DSR}
	\text{\DSR} = \frac{\text{Number of dynamic scatterers}}{\text{Total number of scatterers}}.
\end{equation}
Simulations were conducted for 100 equally distributed \DSRs in the range [0.0, 1.0].
The scatterer density was set to the same as that of Study 1, i.e., 0.055 scatterers/\cbum.
The total number of scatterers and the scatterer density were held constant regardless of the \DSR.

The \DSR dependency of aLIV, Swiftness, LIV, and OCDS was investigated for dynamic scatterer speeds of 0.01, 0.05, 0.1, 0.2, 0.6, and 3 \um/s.
Similar to Study 1, five simulation trials were conducted for each \DSR, and the median among the five trials was taken as the result.

\subsection{Results of numerical validation}
\label{sec:numericalResults}
\subsubsection{Study 1: Scatterer-speed dependency}
\label{sec:numericalResultsSpeed}
\begin{figure}
	\centering\includegraphics{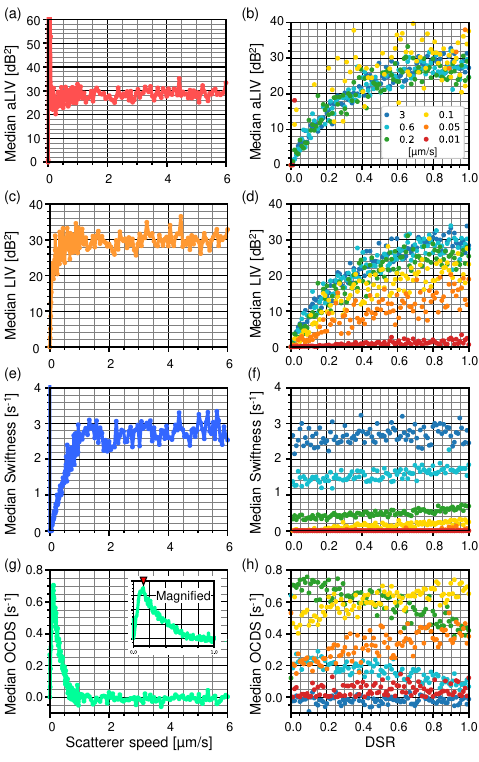}
	\caption{The scatterer-speed dependencies (a, c, e, g) and dynamic-scatterer-ratio (\DSR) dependencies (b, d, f, h) of aLIV, LIV, Swiftness, and OCDS obtained by numerical simulations.
		The point colors of the \DSR-dependency plots indicate the speed of the dynamic scatterers as shown in the legend in (b).
		In the magnified plot for the speed range [0, 1 \um/s] (g), the peak location of OCDS is indicated with the arrowhead.
	}
	\label{fig:simulation results}
\end{figure}
Figure \ref{fig:simulation results}(a, c, e, g) shows the scatterer-speed dependency of aLIV, LIV, Swiftness, and OCDS.

Both aLIV and LIV remain constant at around 30 \sqdb for higher scatterer speeds [Fig.\@ \ref{fig:simulation results}(a, c)], which is consistent with the maximum LIV values reported in previous \exvivo and \invitro studies \cite{Mukherjee2021SciRep, Mukherjee2022BOE, Morishita2023BOE, El-Sadek2024SciRep}.
For slower scatterer speeds, where $|v|$ < 0.2 \um/s, LIV becomes smaller as the scatterer speed decreases.
In contrast, aLIV becomes extremely large and extends beyond the plot region.
This is because the LIV curve cannot be correctly fitted for very slow scatterer speeds.
This issue is extensively discussed in Section \ref{sec:ambiquousLivCurve}. 

Swiftness exhibits saturation with respect to the scatterer speed [Fig.\@ \ref{fig:simulation results}(e)], except for some minor oscillations.
Namely, Swiftness increases monotonically as the scatterer speed increases up to around 1.2 \um/s, and then remains at around 2.5 \invs at higher scatterer speeds.
Thus, Swiftness can be considered to be sensitive to scatterer speeds if it is less than 1.2 \um/s.

OCDS exhibits a pinnacle that peaks around 0.1 \um/s [Fig.\@ \ref{fig:simulation results}(g), magnified in the inset].
This indicates that OCDS is sensitive to a specific speed range.
This is advantageous because it allows OCDS to be used as a fingerprint for detecting a particular speed.
It is noteworthy that the fingerprint speed zone can be customized by adjusting the delay-time range used to fit the slope of the auto-correlation function.

The minor oscillations observed in Swiftness and other unexplained observations will be examined as part of an ongoing comprehensive numerical investigation of DOCT \cite{Feng2025BiOS, Fujimura2025BiOS}.

\subsubsection{Study 2: \DSR dependency}
Figure \ref{fig:simulation results}(b, d, f, h) shows the dependency of aLIV, LIV, Swiftness, and OCDS on \DSR.

Both aLIV and LIV increase monotonically with increasing \DSR [Fig.\@ \ref{fig:simulation results}(b, d)] when the scatterer speed is greater than 0.2 \um/s.
However, when the speed of the dynamic scatterers is less than 0.1 \um/s, aLIV diverges to extremely large values.
For speeds slower than 0.05 \um/s, almost all aLIV values were plotted outside of the plot region.
In contrast, LIV becomes smaller with decreasing scatterer speed, as shown in Fig.\@ \ref{fig:simulation results}(d).

This property of LIV makes it difficult to estimate the \DSR from the LIV value, namely, one value of LIV can correspond to multiple \DSRs at different scatterer speeds.
In contrast, aLIV becomes extremely large and is unreliable when the scatterer speed is slow.
The unreliable values of aLIV can be automatically detected and removed, as discussed in Section \ref{sec:ambiquousLivCurve}.
For cases where aLIV is not divergent, nearly identical aLIV-\DSR curves are observed for all dynamic scatterer speeds [Fig.\@ \ref{fig:simulation results}(b)], indicating a one-to-one relationship between aLIV and \DSR.
Hence, the aLIV value is specific to a certain \DSR value in the range where the aLIV value is reliable.

Swiftness shows a slight increase with increasing \DSR [Fig.\@ \ref{fig:simulation results}(f)], but the impact of \DSR on Swiftness is significantly smaller than that of the scatterer speed.

OCDS shows either a monotonic increase or decrease with respect to changes in \DSR.
A monotonic increase occurs [warm-color plots, Fig.\@ \ref{fig:simulation results}(h)] when the dynamic scatterer speed is slower than the fingerprint peak speed observed in Fig.\@ \ref{fig:simulation results}(g).
Conversely, a monotonic decrease occurs [cool-color plots, Fig.\@ \ref{fig:simulation results}(h)] when the dynamic scatterer speed is faster than the fingerprint peak speed.
The slope of OCDS with respect to \DSR becomes steeper as the dynamic scatterer speed moves closer to the finger print peak in Fig.\@ \ref{fig:simulation results}(g).
However, the change in OCDS with respect to \DSR in Fig.\@ \ref{fig:simulation results}(h) is much smaller than the response to the fingerprint speed in Fig.\@ \ref{fig:simulation results}(g).
Therefore, the conclusion of Study 1, namely, that OCDS can be used as a fingerprint for a particular speed, is still supported.

\subsection{Summary of DOCT metrics' properties}
\label{sec:simulationSummary}
The numerical simulations have demonstrated that aLIV is sensitive to the \DSR, such that aLIV becomes larger for higher values of \DSR,  in the range where the aLIV value is reliable.
In addition, unreliable aLIV values can be automatically detected (Section \ref{sec:ambiquousLivCurve}).
Therefore, aLIV can be used as an indicator of the dynamic scatterer ratio (i.e., the occupancy of the dynamic scatterers).

LIV also exhibits sensitivity to \DSR, similar to aLIV.
However, as shown in Fig.\@ \ref{fig:simulation results}(d), it is difficult to determine a unique \DSR from the LIV value if the scatterer speed is unknown.

Swiftness increases almost monotonically as the scatterer speed increases, especially for speeds of less than 1.2 \um/s.
In addition, Swiftness is relatively insensitive to \DSR.
Thus, Swiftness can be used as an indicator for the speed of dynamic scatterers.
It is noteworthy that, although the monotonic relation can be found only up to 1.2 \um/s, this maximum speed of the monotonic relation can be controlled by the OCT system speed and the scanning protocol.
Specifically, higher speeds can be measured by using a shorter interval of OCT frame acquisition.

OCDS displays a distinct peak at a particular scatterer speed, indicating that this metric can be used as a fingerprint for a specific scatterer speed.
Namely, the high-value of OCDS indicates the scatterer motion at a specific speed range. 
The principle of OCDS suggests that the fingerprint speed zone can be selected by changing the delay-time range.
Namely, if we double the inter-frame time for example, the finger-print speed also doubles.
Furthermore, the location and width of the fingerprint speed zone can be identified from numerical simulations.
It should be noted that, in contrast to the markedness of the high OCDS value, the low OCDS cannot discriminate whether the scatterer speed is slower or faster than the specific range.

\section{Experimental validation of new DOCT metrics}
\label{sec:experimental validation}
To assess the applicability of aLIV and Swiftness to biological samples, we imaged  \invitro tumor spheroids and \exvivo mouse kidney samples.
\subsection{Method of experimental validation}
\subsubsection{Samples}
Four tumor spheroids treated with an anti-cancer drug were involved.
The tumor spheroids were formed with human breast cancer cells (MCF-7 cell line) in each well of a 96-well plate under a cultivation environment.
Three of the four spheroids were treated with 1-\uM paclitaxel (PTX) for 1, 3, and 6 days, respectively, while the remaining spheroid was untreated.

After drug treatment, each spheroid was measured using an OCT microscope.
The cultivation and measurement protocols have been detailed elsewhere in Ref.\@ \cite{El-Sadek2023SciRep}, and the raw data used in the present study are identical to those presented in this reference.

An \exvivo C57BL/6 healthy mouse kidney was also examined in this study.
The sample preparation has been described elsewhere in Ref.\@ \cite{Mukherjee2023SciRep}, and the raw OCT data presented in Fig.\@ 1 of this reference was used in this study.

The mouse kidney experiments were performed in accordance with the animal study guidelines of the University of Tsukuba.
All experimental protocols were approved by the Institutional Animal Care and Use Committee (IACUC) of the University of Tsukuba.
The present study was designed, performed, and reported according to Animal Research: Reporting of \Invivo Experiments (ARRIVE) guidelines.
The raw OCT data of Ref.\@ \cite{Mukherjee2023SciRep} was used, and thus no additional animal experiments were conducted for the present validation.

\subsubsection{OCT microscope and DOCT measurement}
\label{subsubsec:OCT and DOCT}
A custom-made Jones-matrix swept-source OCT microscope with a scan speed of 50,000 A-lines/s was used for the measurements.
This OCT microscope has a probe beam center wavelength of 1.3 \um and resolutions of 18 \um ($1/e^2$ width) and 14 \um (FWHM in tissue) for the lateral and axial directions, respectively.
Although this OCT microscope is polarization sensitive, the DOCT imaging did not use polarization information.
The polarization-insensitive OCT intensity signal was obtained by averaging linear intensities of four OCT images acquired from four different polarization channels.
The OCT system is described in detail elsewhere \cite{Li2017BOE, Miyazawa2019BOE}.

Time-sequential OCT data were acquired using a repeating raster scan protocol \cite{El-Sadek2021BOE}.
The transversal fields of view (FOVs) were 1 mm $\times$ 1 mm for the tumor spheroids and 6 mm $\times$ 6 mm for the mouse kidney.
The \enface FOV was divided into eight sub-fields along the slow scan direction, with each sub-field consisting of 16 B-scan locations.
The raster-scan was performed 32 times for each sub-field.
As a result, a time series of 32 frames was acquired at each B-scan location.
The inter-frame time was 204.8 ms and the time separation between the first and last frames was 6.35 s.
Each frame comprised 512 A-lines.

For DOCT imaging, aLIV and Swiftness were computed as described in Section \ref{sec:principle}.
LIV and OCDS were computed by the method described in Section \ref{subsubsec: simulation_DOCT method}.
We applied 3$\times$3-pixel kernel averaging in the cross-sectional plane to each of the LIV and OCDS images.
The pseudo-color DOCT images were generated using the OCT intensity and the DOCT as the pixel brightness and hue, respectively.

\subsubsection{Study design and protocol}
We conducted two studies.
In Study 1, the image appearances were subjectively compared between four types of DOCT images (i.e., aLIV, Swiftness, LIV, and OCDS).
We also used fluorescence images for the biological interpretation of the tumor spheroids.

In Study 2, we investigated the impact of $\Atw$ on the DOCT metrics and the image appearances.
It is because low dependency of DOCT metrics on $\Atw$ can be beneficial for future applications that require shorter acquisition times, such as drug screening by large numbers of \invitro samples and \invivo imaging.
Shorter acquisition times enable longitudinal assessment of \invitro samples at fine measurement time intervals and help reduce the effects of bulk motion during \invivo imaging.

To examine the effect of $\Atw$, we virtually shortened $\Atw$ by truncating the frame sequence, which initially contained 32 frames.
Here, the virtually shortened $\Atw$ is referred to as the virtual acquisition time window ($\VAtw$).
Six different $\VAtw$ were investigated, i.e., $\VAtw$ = 0.41, 1.23, 2.46, 3.69, 4.91, and 6.35 s.
The time-sequential OCT data for each $\VAtw$ consisted of a different number of frames, i.e., 3, 7, 13, 19, 25, and 32 frames.

Quantitative metrics-for-comparison were calculated from two regions of interest (ROIs) in the untreated spheroid.
The ROIs were manually defined as shown in Fig.\@ \ref{fig:result2_quantitativeAnalysis}(e).
It is known that the spheroids have two domains, a necrotic core and a vital periphery.
Each domain exhibits different DOCT values and is mostly homogeneous.
Thus, each ROI was defined in core and periphery regions. 

We computed two types of metrics including the average DOCT value at each ROI, and the contrast between the core and periphery ROIs defined as 
\begin{equation}
	\label{eq:contrast}
	C = \frac{|M_c - M_p|}{M_c + M_p},
\end{equation}
where $M_c$ and $M_p$ represent the average values in the core and periphery ROIs, respectively.
Some pixels produced unreliable aLIV and Swiftness values because of the low fitting accuracy of the LIV curve.
These pixels were detected automatically and excluded from the averaging.
Details of the unreliable pixels and their automatic detection are described in Section \ref{sec:ambiquousLivCurve}.

\revstart{c2-5}
It should be noted that the protocol of Study 2 is suboptimal, as the number of frames varies among the $\VAtw$ configurations.
However, the effect of varying frame numbers is expected to be small for the following reasons.
Specifically, all four types of DOCT metrics are derived from the variance of OCT signals.
OCDS, in particular, is calculated from the correlation coefficient, which is defined using variances and covariance.
As is well known in statistical theory, the stability of the variance estimate decreases as the number of data points decreases.
This reduction in stability will reduce the overall image quality of the DOCT images.
On the other hand, the expectation of variance is not significantly affected by the number of data points.
\revend

The mouse kidney was not used in Study 2, because it did not contain simple large-domain structures.

\subsection{Results}
\subsubsection{Study 1}
\label{sec:resultStudy1}
\begin{figure}
	\centering\includegraphics{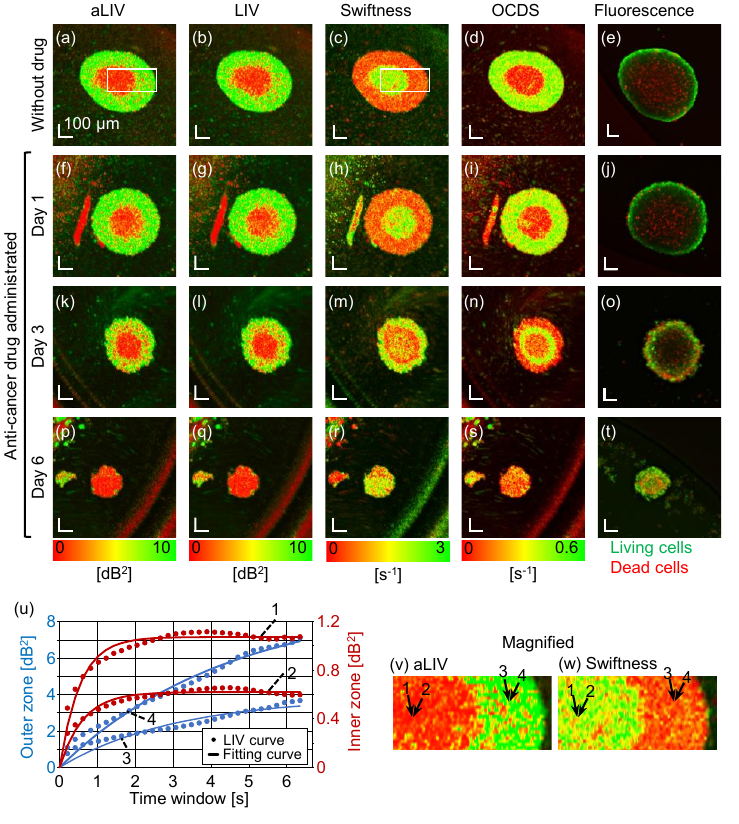}
	\caption{The \enface aLIV, LIV, Swiftness, OCDS, and fluorescence images of the tumor spheroids treated by the anti-cancer drug (1-\uM PTX) for 1, 3, and 6 days.
		The scale bars represent 100 \um.
		In the region with high aLIV (green), the dynamic scatterer ratio is expected to be high, while in the region with high Swiftness, the dynamic scatterers are expected to move rapidly.
		\revstart{c2-1}
		Examples of LIV curves are shown in (u), which were extracted at the points indicated by the arrows in (v, w).
		(v, w) correspond to the square regions in (a, c).
		\revend
	}
	\label{fig:result1a_spheroid}
\end{figure}
Figure \ref{fig:result1a_spheroid} presents the \enface aLIV, LIV, Swiftness, OCDS, and fluorescence images (from left to right) of the untreated (1st row) and anti-cancer (1-\uM PTX) drug-treated tumor spheroids (treatment for 1, 3, and 6 days shown in 2nd, 3rd, and 4th rows, respectively).
\revstart{c2-9}
The fluorescence images were captured using a wide-field non-confocal fluorescence microscope (THUNDER imager DMi8; Leica Micro-systems, Wetzer, Germany) with a microscopic objective lens having a numerical aperture (NA) of 0.12.
Living cells were highlighted with calcein-acetoxymethyl (calcein-AM; Dojindo, Kumamoto, Japan), which emits a green fluorescence signal, while dead cells were highlighted with propidium iodide (PI; Dojindo), which emits a red fluorescence signal.
\revstart{c2-9}
Since the NA of the fluorescence imaging is low, the fluorescence image is not really depth sectioned.
This should be taken into account when the \enface slice of DOCT and the fluorescence image are compared. 
\revend

For the without-drug and 1-day-treated spheroids, a concentric pattern consisting of two zones is observed in both the DOCT and fluorescence images [Fig.\@ \ref{fig:result1a_spheroid}(a-j)].
The fluorescence images show dead cells (red) concentrated in the inner zone, while living cells (green) are distributed in the periphery. 
These characteristics are consistent with the well-known necrotic core formation of MCF-7 spheroids \cite{Costa2016BA}, which is caused by hypoxia and nutrient deficiency in the central region.
Therefore, the inner and outer zones of the concentric pattern in the DOCT images may also correspond to the necrotic core and the living cells, respectively.

\revstart{c2-1}
Four representative LIV curves (curves 1-4) and their fitting curves are shown in Fig.\@ \ref{fig:result1a_spheroid}(u).
The curves 1 and 2 are obtained at the inner zone, while curves 3 and 4 are obtained at the outer zone as indicated in Fig.\@ \ref{fig:result1a_spheroid}(v, w).
These two zones exhibited distinct characteristics in their LIV curves.
Specifically, the outer zone [blue plots in Fig.\@ \ref{fig:result1a_spheroid}(u)] shows the LIV curves with a larger magnitude than the inner zone (red plots), while the inner zone shows the LIV curve with a faster saturation than the outer zone
\revend

The inner zone, i.e., necrotic core, exhibits low aLIV, low LIV, high Swiftness, and low OCDS.
According to the simulation results, the DOCT contrast indicates that only a small proportion of scatterers within the necrotic core are dynamic (as suggested by the low aLIV and LIV), but these dynamic scatterers move rapidly (as suggested by the high Swiftness).
The speed of the dynamic scatterers is outside the sensitive range, i.e., fingerprint speed zone, of OCDS (as suggested by the low OCDS, see Section \ref{sec:numericalResultsSpeed}).
The Swiftness values further confirm that the speed exceeds the fingerprint speed zone of OCDS.
In contrast, the outer zone, i.e., spheroid periphery, contains a large proportion of dynamic scatterers (high aLIV and LIV), but these dynamic scatterers move slowly (low Swiftness) and their speed is within the fingerprint speed zone of OCDS, resulting in the high OCDS value.

For the 3-day-treated spheroid, the Swiftness and OCDS images reveal a clear three-zone concentric pattern [Fig.\@ \ref{fig:result1a_spheroid}(m, n)].
Similarly, aLIV and LIV exhibit a concentric pattern of low (red)--moderate (yellow with red dots)--high (green) zones from the inner to outer regions [Fig.\@ \ref{fig:result1a_spheroid}(k, l)].
In the innermost zone, a very small proportion of scatterers are dynamic (low aLIV and LIV) and move at high speed (high Swiftness).  
In the middle zone, both static and dynamic scatterers coexist (moderate aLIV and LIV), with dynamic scatterers moving at slower speeds (low Swiftness) within the fingerprint speed zone of OCDS (high OCDS).
The outermost zone of the aLIV and LIV images resembles the outer zone of the without-drug and 1-day-treated spheroids, namely, a large proportion of the scatterers are dynamic (high aLIV and LIV).
However, Swiftness and OCDS exhibit reverse contrasts to those of the without-drug and 1-day-treated spheroids.
This indicates that the dynamic scatterers move rapidly (high Swiftness) and their speed is outside the fingerprint speed zone of OCDS (low OCDS).
It is noteworthy that the three concentric zones were not visible in the fluorescence image, which visualizes a mixture of live and dead cells.

For the 6-day-treated spheroid, aLIV and LIV indicate that a very small proportion of scatterers are dynamic (low aLIV and LIV) across the entire spheroid [Fig.\@ \ref{fig:result1a_spheroid}(p, q)].
However, Swiftness reveals a tessellated pattern, with some domains containing fast-moving scatterers and others containing slow-moving scatterers [Fig.\@ \ref{fig:result1a_spheroid}(r)].
A similar tessellated pattern is evident in the OCDS image [Fig.\@ \ref{fig:result1a_spheroid}(s)].
The fluorescence image also shows a similar tessellated pattern of dead and living cell domains [Fig.\@ \ref{fig:result1a_spheroid}(t)].

\begin{figure}
	\centering\includegraphics{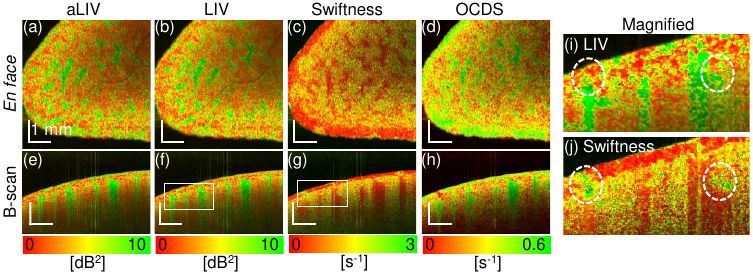}
	\caption{The \enface and B-scan aLIV, LIV, Swiftness, and OCDS images of the \exvivo mouse kidney.
		The scale bars represent 1 mm.
		The magnified images of LIV and Swiftness (i, j) correspond to the square regions in the B-scan images (f, g), respectively.
		Several pipe-like structures are visible, and they exhibit high aLIV values, which may indicate high occupancy of dynamic scatterers (high dynamic scatterer ratio), and low Swiftness values, suggesting the motion of the dynamic scatterers is slow.
		}
	\label{fig:result1b_kidney}
\end{figure}
Figure \ref{fig:result1b_kidney} shows the \enface and B-scan aLIV, LIV, Swiftness, and OCDS images of the \exvivo mouse kidney.
Similar to the characteristics reported by Mukherjee \etal \cite{Mukherjee2023SciRep}, pipe-like structures with a large proportion of dynamic scatterers (high aLIV and LIV) can be observed in the inner region of the kidney in the \enface images [Fig.\@ \ref{fig:result1b_kidney}(a, b)].
The low Swiftness and low OCDS values indicate that the dynamic scatterers in these structure move slowly (low Swiftness) and their speed is within the fingerprint speed zone of OCDS (high OCDS) [Fig.\@ \ref{fig:result1b_kidney}(c, d)].

Vertical stripes appear in the B-scan images [Fig.\@ \ref{fig:result1b_kidney}(e--h)]. 
Mukherjee \etal investigated the LIV of mouse kidneys, and hypothesized that the vertical stripes of high LIV are artifacts caused by fast motion at the upper-most part of the stripes, and hence this vertical stripe is similar to projection artifacts observed in OCT angiography of the retina \cite{Mukherjee2023SciRep, Spaide2015Retina}.
However, this hypothesis was not confirmed in that study because only LIV was used for DOCT imaging in Ref.\@ \cite{Mukherjee2023SciRep}.
On the other hand, the Swiftness image obtained in the present study provides supporting evidence.
Namely, at the top of the vertical stripes [dotted circles in Fig.\@ \ref{fig:result1b_kidney}(i, j), which are magnified images of the square regions in Fig.\@ \ref{fig:result1b_kidney}(f, g)], high Swiftness values are observed, indicating high-speed dynamic scatterers.
These regions exhibit low OCDS, probably because the scatterer speeds exceed the fingerprint speed zone of OCDS.

\revstart{c2-4}
It should be noted that the interpretations according to the simulation results rely on several assumptions, such as that a single motion model (i.e., random ballistic motion model) and a single scatterer speed govern the motion.
The scatterer speed used in the simulation is confined to a reasonable range for the scatterers in the samples.
These limitations are discussed in detail in the discussion section (Section \ref{sec:discuss_complicateModel}).
\revend
\subsubsection{Study 2}
\begin{figure}
	\centering\includegraphics{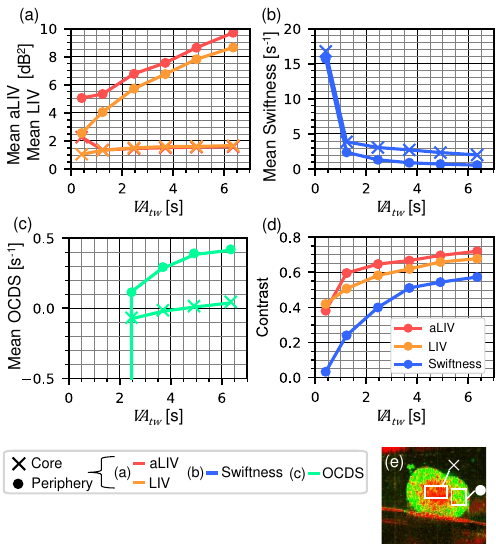}
	\caption{(a, b, c) Mean DOCT values in each core ($\times$) and periphery ($\medbullet$) ROI of the spheroids.
		(d) Contrast between core and periphery ROIs.
		The two ROIs were manually defined as shown by the squares in (e).
		}
	\label{fig:result2_quantitativeAnalysis}
\end{figure}
Figure \ref{fig:result2_quantitativeAnalysis}(a--d) presents the $\VAtw$ (virtual acquisition time window) dependency of each DOCT metric.
Specifically, Fig.\@ \ref{fig:result2_quantitativeAnalysis}(a--c) shows the mean DOCT values of aLIV and LIV [Fig.\@ \ref{fig:result2_quantitativeAnalysis}(a)], Swiftness [Fig.\@ \ref{fig:result2_quantitativeAnalysis}(b)], and OCDS [Fig.\@ \ref{fig:result2_quantitativeAnalysis}(c)] at the two ROIs.
The core and periphery ROIs used for the analysis are indicated in Fig.\@ \ref{fig:result2_quantitativeAnalysis}(e).

Here we note that the DOCT values might be more faithful with larger $\VAtw$ values because the larger $\VAtw$ correspond to longer acquisition time duration and greater number of time points.
All DOCT metrics exhibit a clear dependency on $\VAtw$.
Specifically, as $\VAtw$ becomes smaller, aLIV, LIV, and OCDS decrease while Swiftness increases. 

The contrast of each DOCT metric between the core and periphery ROIs is plotted in Fig.\@ \ref{fig:result2_quantitativeAnalysis}(d).
The OCDS of the core ROI exhibits erroneous negative values for small $\VAtw$ values, thus, the contrast of the OCDS was not computed.
These erroneous values may be caused by the limited number of time points involved in the OCDS computation.
For aLIV, LIV, and Swiftness, the contrast becomes large and stable as $\VAtw$ increases.
It appears that aLIV stabilizes faster than LIV.
Considering that the DOCT values are more reliable at larger $\VAtw$, aLIV might be a more reliable metric than LIV.

The contrast between the core and periphery ROIs for aLIV, LIV, and Swiftness decreases monotonically as $\VAtw$ becomes smaller [Fig.\@ \ref{fig:result2_quantitativeAnalysis} (d)].
The aLIV metric demonstrates higher contrast than LIV, except at $\VAtw$ = 0.41 s.
This higher contrast of aLIV agrees with our expectations based on the principle of aLIV.
Namely, the saturation speed of the LIV curve is slow and it does not reach the saturation level within $\Atw$, such as in the case of the periphery ROI, the aLIV value, which is the expected saturation level of the curve, becomes larger than the LIV values.
On the other hand, if the LIV curve saturates fast, such as in the case of core ROI, the aLIV and LIV values become close to each other.
And hence, the contrast of aLIV between two ROIs can be higher than that of LIV, especially $\Atw$ is small.
Note that the case of the shortest $\Atw$ (0.41 s) is an exception, where the $\Atw$ is too short to correctly estimate the saturation level (that is the aLIV).

\begin{figure}
	\centering\includegraphics{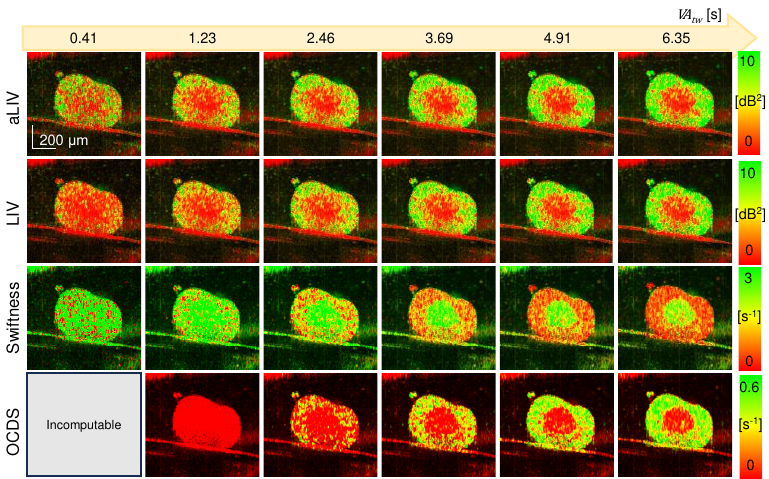}
	\caption{Image quality comparison of aLIV, LIV, Swiftness, and OCDS for several $\VAtw$ values.
		The original data is that of the without-drug spheroid in Fig.\@ \ref{fig:result1a_spheroid}.
	}
	\label{fig:result2_spheroid}
\end{figure}
Figure \ref{fig:result2_spheroid} shows B-scan images of aLIV, LIV, Swiftness, and OCDS of a tumor spheroid at each $\VAtw$.
The observational contrasts are consistent with the contrast plot of Fig.\@ \ref{fig:result2_quantitativeAnalysis}.
The image quality of all DOCT metrics degrades as $\VAtw$ decreases.
According to subjective observations, the image quality of aLIV and LIV with $\VAtw$ = 3.69 s or larger is acceptable.
In the Swiftness image, the spheroid core is distinguishable for $\VAtw$ = 1.23 s or larger, while for OCDS, the core is distinguishable for $\VAtw$ = 2.46 s or larger.
\revstart{c2-5}
The degradation of the image quality at regions with small $\VAtw$ values might be caused by the instability of the variance estimate due to the reduced number of data points.
\revend

\begin{figure}
	\centering\includegraphics{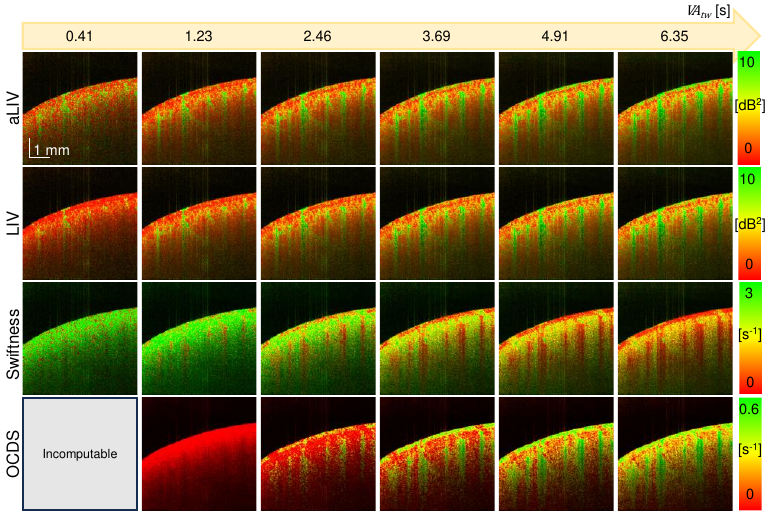}
	\caption{Image quality comparison of aLIV, LIV, Swiftness, and OCDS for each $\VAtw$.
		The original data is that of the mouse kidney in Fig.\@ \ref{fig:result1b_kidney}.
	}
	\label{fig:result2_kidney}
\end{figure}
Figure \ref{fig:result2_kidney} shows B-scan images of aLIV, LIV, Swiftness, and OCDS for the mouse kidney at each $\VAtw$.
The DOCT images of the mouse kidney exhibit a similar $\VAtw$ dependency as observed for the spheroid.

\section{Discussion}
\subsection{DOCT signals and intratissue/intracellular motions}
Most of the DOCT algorithms are not direct measures of sample dynamics (i.e., intratissue and intracellular motions) but measures of OCT signal fluctuations.
In general, complex OCT signal is expressed as a summation of phasor contributions from scatterers in the tissue.
\revstart{c1-1}
The time frequency of the complex OCT sequence, e.g., obtained by a Fourier-spectrum-based DOCT method \cite{SchulzHildebrandt2024FMN}, may directly correspond to the vibration of the phasors, but the vibration of a phasor does not necessarily indicate the vibration of the scatterers.
\revend
It can be easily imagined that the phase of the phasor rotates even when the scatterer moves with constant linear motion.
Furthermore, several Fourier-spectrum-based DOCT methods use the time-frequency spectrum of the intensity or amplitude OCT signal sequence \cite{Ling2017LSM, Thouvenin2017IOVS, Munter2020OL, Leung2020BOE, Xia2023Optica, KYChen2024BOE}.
According to discrete-scatterer-based formulations of OCT \cite{Zhou2021AOP, Tomita2023BOE}, the intensity of OCT consists of interaction terms (i.e., the multiplications) of the phasors (see Section 3.2 and Fig.\@ 2 of Ref.\@ \cite{Tomita2023BOE}).
These interactions generate new frequency components that were not in the original complex OCT sequence.
And hence, the Fourier-spectrum-based DOCT contrasts are not really direct measures of the sample motion.
Note that, since OCT amplitude is defined as the square root of intensity, amplitude-based methods are also not free from this issue.
(Note that the Taylor series expansion of the square root may have infinitely high order terms.)

Similarly, autocorrelation-based methods \cite{Oldenburg2015Optica, Apelian2016BOE, El-Sadek2020BOE} are also not really direct measures of sample dynamics.
It can be understood from the fact that the autocorrelation function and the power spectrum are tightly related by Wiener-Khinchin theorem.
Note that, among the autocorrelation-based methods, OCDS \cite{El-Sadek2020BOE} uses logarithmic intensity, and the Taylor series expansion of the logarithmic function also consists infinitely high order terms.
In the other words, while logarithmic scaling helps reduce sensitivity to overall optical power fluctuations, it also amplifies local fluctuations and noise.
And hence, OCDS is also not a direct measure of the sample dynamics.

Our new DOCT metrics, i.e., aLIV and Swiftness, are also not free from this issue.
However, the numerical analyses revealed relatively straightforward relationships between these metrics and the sample dynamics (Section \ref{sec:numerical simulation}).
Namely, aLIV mainly corresponds to DSR, while Swiftness mainly corresponds to the scatterer speed.
These simple relationships enable easy interpretation of DOCT images.
\revstart{c2-8}
In addition, since the aLIV-DSR curves are nearly identical among all scatterer speeds, aLIV provides a more reliable estimate of the DSR than conventional LIV. This advantage is especially effective when analyzing real samples with a wide distribution of scatterer speeds. 
\revend
In addition, these properties of aLIV and Swiftness will also enable quantitative measurement of DSR and scatterer speed by combining with a sophisticated estimation theory as shown in a preliminary demonstration\cite{Fujimura2025BiOS}.

\subsection{Ambiguous LIV curve and unreliable aLIV and Swiftness values}
\label{sec:ambiquousLivCurve}
\revstart{c2-3}
\begin{figure}
	\centering\includegraphics{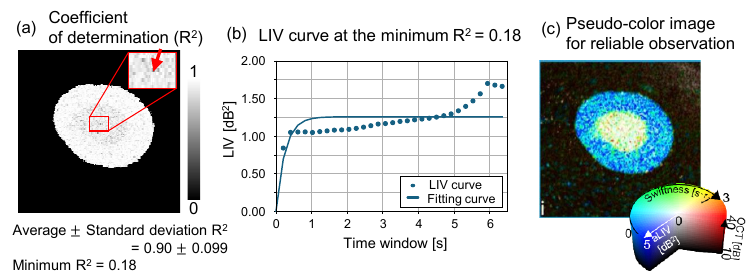}
	\caption{(a) Coefficient of determination (R$^2$ value) map of the without-drug spheroid data.
		The region outside the tissue was masked out.
		(b) LIV curve at the pixel exhibiting the minimum R$^2$ value, indicated by an arrow in (a).
		(c) Example of the pseudo-color image method, reprinted from Ref.\@ \cite{Noah2025BOE}.
	}
	\label{fig:fittingAccuracy}
\end{figure}
The fitting accuracy of the LIV curve was investigated by calculating the coefficient of determination (R$^2$ value) for the without-drug spheroid data as shown in Fig.\@ \ref{fig:fittingAccuracy}(a).
R$^2$ ranges from 0 for the worst fit to 1 for the best.
There are some pixels with small R$^2$ values.
For example, the LIV curve with the smallest R$^2$ (R$^2$ = 0.18) (at the location indicated by the arrow) cannot be properly represented by the saturation function [Fig.\@ \ref{fig:fittingAccuracy}(b)].
Specifically, the LIV is nearly saturated at the first time-window and exhibits a slight subsequent increase.
Since the LIV values on this curve is very small, we suspect that this slight subsequent increase is mainly caused by noise rather than intratissue scatterer motions.
Although a few pixels show low  R$^2$ values, a good mean R$^2$ value was observed within the tissue as  0.90 $\pm$ 0.099 (mean $\pm$ standard deviation).
This indicates that most LIV curve fittings were performed with reasonable fitting accuracy.

However, the good fit (i.e., large R$^2$) does not always ensure the good stability of the fitting parameters (i.e., saturation level and time constant).
\revend
There are two types of LIV curves that inevitably lead to unreliable parameter fitting, and hence produce unreliable aLIV and Swiftness values.

The first such LIV curve, referred to as Type-A ambiguous LIV curve, constantly takes values close to 0 \sqdb regardless of $\tw$ (the time window of the variance computation).
This type of LIV curve corresponds to the case in which there are almost no dynamic scatterers.
In this case, myriad $(a, \tau)$-pairs of Eq.\@ (\ref{eq:1st-order saturation function}) provide a solution to the curve fitting.
Namely, if $a$ is zero, any value of $\tau$ gives a solution.
Similarly, if $\tau$ approaches $\infty$, any value of $a$ can be a solution.
Hence, aLIV and Swiftness are not definitive with this type of LIV curve.

The second unreliable LIV curve, referred to as Type-B ambiguous LIV curve, is characterized by its monotonic increase with respect to $\tw$ without becoming saturated, namely, the true $\tau$ (i.e., the true time constant) is significantly larger than the maximum acquisition time $\Atw$.
This type of LIV curve corresponds to the case in which the dynamic scatterers move very slowly.
In this case, the fitting algorithm needs to predict the saturation level $a$ that the curve may reach outside of the fitting range.
Namely, the algorithm needs to extrapolate the data sequence, which causes instability and inaccuracy of fitting, and hence, produce unreliable aLIV and Swiftness values.

Although these two types of LIV curves cause unreliable aLIV and Swiftness computations, they can be automatically detected and removed from quantitative analyses.
The Type-A ambiguous LIV curve is characterized by persistent small values around zero, and hence can be detected by the following criterion:
\begin{equation}
	\label{eq:curveADetection}
	\sqrt{\left\langle\overline{\LIV}^2(\tw)\right\rangle_{\tw}} 
	\leq 2 \sqrt{\left\langle\left(\overline{\LIV}(\tw) - f_\LIV(\tw; a, \tau)\right)^2\right\rangle_{\tw}},
\end{equation}
where $\langle\quad\rangle_{\tw}$ represents averaging over $\tw$, $\overline{\LIV}(\tw)$ is the LIV curve, and $f_\LIV(\tw; a, \tau)$ is the estimated first-order saturation function.
Namely, if the root-mean-square of the LIV curve is less than twice the root-mean-square-error of the fitting, $\overline{\LIV}(\tw)$ is considered to be constantly zero and the curve is classified to be a Type-A ambiguous LIV curve.

The Type-B ambiguous LIV curve corresponds to the case in which the time constant $\tau$ is far larger than the maximum acquisition time $\Atw$.
This can be detected by the following criterion:
\begin{equation}
	\label{eq:curveBDetection}
	\tau \geq 2 \Atw.
\end{equation}
Namely, if the estimated time constant $\tau$ is greater than twice the maximum acquisition time, the curve is considered to be a Type-B ambiguous LIV curve.

Occurrences of the two types of ambiguous LIV curves were examined using the experimentally obtained datasets of Section \ref{sec:experimental validation}.
The ambiguous LIV curves were detected by Eqs.\@ (\ref{eq:curveADetection}) and  (\ref{eq:curveBDetection}), and the occurrences were computed within the tissue region selected by an empirical intensity threshold.
Table \ref{tab:ambiguousCurve} summarizes the mean and standard deviation of the occurrences among the five datasets (i.e., four spheroids and mouse kidney).
The numbers in the table denote the percentage of curves (i.e., pixels in the aLIV and Swiftness images) of each types.
On average, more than 97\% of the LIV curves are classified as unambiguous (i.e., neither Type-A nor Type-B).
Namely, more than 97\% of the pixels in the aLIV and Swiftness images are unambiguous.
\begin{table}[ht]
	\caption{Occurrence of ambiguous LIV curves with in the tissue region.
		Each cell indicates the mean and standard deviation of the percent-occurrences among five samples.
		The results indicate that more than 97\% of pixels in the tissue regions are unambiguous.
	}
	\label{tab:ambiguousCurve}
	\centering\includegraphics{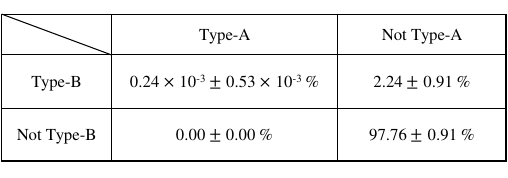}
\end{table}

The impact of exclusion of the unreliable pixels caused by ambiguous LIV curves is now examined using the spheroid dataset without drug administration.
The means and standard deviations of aLIV and Swiftness in the core and periphery ROIs were computed with (w/) and without (w/o) unreliable pixels, and the results are summarized in Table \ref{tab:unreliableExclusion}.
\begin{table}[ht]
	\caption{Mean $\pm$ standard deviation of aLIV and Swiftness values at two ROIs with (w/) and without (w/o) unreliable pixels.
	}
	\centering\includegraphics{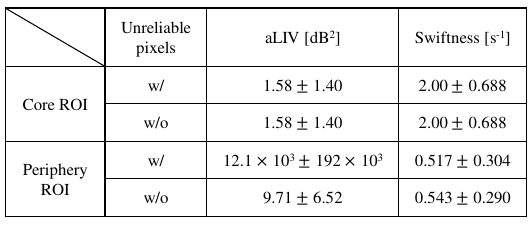}
	\label{tab:unreliableExclusion}
\end{table}

The exclusion of unreliable pixels results in a marked reduction in the mean aLIV value in the periphery ROI.
Our specific curve fitting algorithm tends to give very high estimate of $a$ if it is ambiguous, and it causes the very high mean aLIV value.
This result suggests that the exclusion of unreliable pixels is essential when using the DOCT metrics for quantitative analyses.

\revstart{c2-3}
Furthermore, for observation purposes, we can mitigate this issue by introducing a new pseudo-color-image generation method which is presented in Ref.\@ \cite{Noah2025BOE}.
This method assigns Swiftness, aLIV, and OCT intensity to the hue, saturation, and brightness channels, respectively as exemplified in Fig.\@ \ref{fig:fittingAccuracy}(c), where the image was reprinted from Ref.\@ \cite{Noah2025BOE} and is generated from the same dataset with Fig.\@ \ref{fig:fittingAccuracy}(a).
\revend

\subsection{Processing time of aLIV and Swiftness}
\label{subsec:processing time}
\begin{figure}
	\centering\includegraphics{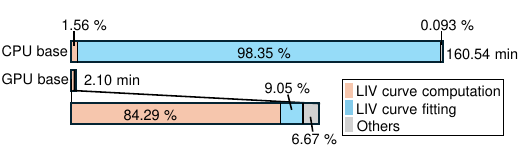}
	\caption{Processing time of aLIV and Swiftness by the CPU- and GPU-based fitting methods.
		The total processing time by the GPU-based method is approximately 2.10 min, which is 76.44 times faster than the CPU-based method.
		The total processing time includes LIV curve computation, LIV curve fitting, and other minor processes.}
	\label{fig:processing time}
\end{figure}
In our DOCT imaging, volumetric time-sequential OCT data were acquired under the following configurations: 128 locations per single volume, 32 frames per B-scan location, and 512 $\times$ 402 pixels per frame.
The total processing time for computing aLIV and Swiftness (excluding standard OCT signal reconstruction) for a volume was approximately 2.10 min in our GPU-based environment (described in Section \ref{subsec:implementation new DOCT}) excluding the hard-disk access time.
As summarized in Fig.\@ \ref{fig:processing time}, the computation of the LIV curve and the LIV curve fitting accounted for 84.29\% and 9.05\% of the total processing time, respectively.

The LIV curve fitting was implemented using the GPU-based fitting method ``gf.fit\_constrained()'' in the Gpufit library \cite{Przybylski2017SciRep}.
To evaluate the necessity for GPU-based processing, the GPU-based fitting method was compared with a CPU-based fitting method, namely ``scipy.optimize.curve\_fit()'' in the SciPy 1.5.2 library.
The model function, algorithm, and parameter exploration range were identical to those in the GPU based method.

The total processing time for aLIV and Swiftness using the CPU-based fitting method was approximately 160.54 min.
The processing times for LIV curve fitting were 0.19 min and 157.89 min for the GPU- and CPU-based fitting methods, respectively.
This verification confirms that the GPU-based fitting method accelerates the LIV curve fitting by a factor of 831.0.

\subsection{Current and future validations of DOCT}
Experimental validation of DOCT methods using dynamic phantoms is important.
However, dynamic phantoms with known and well-controlled dynamics have not yet been established.
Although scatterer suspensions can be used as phantoms to model diffusion motion, diffusion alone does not accurately represent the dominant motion in cells \cite{Feng2025BOE}.
As a result, experimental quantitative validation remains challenging.
Alternatively, we validated the DOCT algorithms through simulations using a numerical sample model, which can be regarded a well-controlled numerical phantom.

Apart from the phantom-based validation, comparing DOCT images with established tissue and cell imaging techniques, such as histology and fluorescence microscopy, is also important.
In this study, we compared DOCT images of spheroids with standard and well-established live/dead fluorescence micrographs (Fig.\@ \ref{fig:result1a_spheroid}), which significantly aided the interpretation of DOCT images.
However, such a comparison has not yet been conducted for \exvivo kidney samples.
Although a detailed interpretation of kidney dynamics is beyond the scope of this paper, such comparisons will contribute to a more comprehensive understanding of DOCT images of the kidney in future studies.

\subsection{Numerical investigation for more complicated scatterer dynamics}
\label{sec:discuss_complicateModel}
In our numerical simulations, we simplified intracellular and intratissue motions using a single motion model, i.e., random ballistic motion model (Section \ref{sec:sampleModel}), although a diffusion (Brownian) motion model can also be considered as the motion model.
A previous study using a numerical simulation revealed that LIV and OCDS exhibit similar characteristics under both the random ballistic and diffusion models \cite{Feng2025BOE}.
Since aLIV and Swiftness are computed based on LIV calculations, we considered that aLIV and Swiftness would show similar results among two motion models if LIV shows similar results.
Therefore, we considered that the fundamental characteristics of DOCT contrasts can be clarified from the results of one of the motion models, namely the random ballistic motion model.

\revstart{c1-4}
However, neither the random ballistic motion model nor the simple diffusion model may represent all types of intratissue activities.
To comprehensively examine all intratissue activities, a more complex model might be necessary.
\revend

Feng \etal expanded the modeling of intratissue scatterer motions by introducing mono-directional displacement (flow) and diffusion (Brownian) motion models, in addition to the random ballistic motion model \cite{Feng2025BOE}.
\revstart{c2-2}
Besides these models, motion models must also account for more ordered or directed intracellular motions, such as cytoplasmic streaming.
However, our current random ballistic motion model is expected to recapitulate such motion only as long as the OCT resolution is larger than the domain size of the motion (such as the cell size).
Specifically, if the resolution volume encompasses multiple cells, and the directional motions within those cells are not correlated to each other, these ordered or directed motions can be considered random motion to some degree.

In any case, a greater variety of motion models is anticipated.
This comprehensive set of motion models could enable the DOCT metrics to be interpreted for a wide spectrum of sample types beyond cellular samples, such as those exhibiting cellular migration and flow, and more aqueous samples.
\revend

\section{Conclusion}
In this paper, we proposed a new DOCT algorithm.
By leveraging the $\Atw$-size dependency of LIV, two DOCT metrics, named aLIV and Swiftness, were derived.
Numerical simulations revealed that aLIV is sensitive to the occupancy of dynamic scatterers (dynamic scatterer ratio), whereas Swiftness is sensitive to the speed of the dynamic scatterers.
Each metric has a one-to-one relationship with either the occupancy or the speed of moving scatterers, respectively.
This indicates that the unique values of occupancy and speed can be determined directly from aLIV and Swiftness.
In addition, we emphasize that these two metrics, which are sensitive to different aspects of sample activity, can be obtained simultaneously.

Experimental validations using \invitro tumor spheroids and an \exvivo mouse kidney demonstrated that aLIV and Swiftness images provide more direct insights into the intratissue activities than our conventional DOCT metrics.
Furthermore, the experimental results suggest that aLIV and Swiftness are relatively insensitive to the acquisition time window in comparison to our conventional DOCT algorithms (LIV and OCDS).
This characteristic might be particularly advantageous for evaluating biological samples requiring short measurement times, such as \invivo samples.

\revstart{c2-4}
It should be noted that our current numerical simulation uses a simplified model for intratissue scatterer motions, which includes several assumptions.
Nevertheless, bolstered by numerical-simulation-based interpretation, the multiple DOCT metrics will enhance the utility and applications of DOCT imaging.
\revend

\section*{Funding}
Core Research for Evolutional Science and Technology (JPMJCR2105); Japan Society for the Promotion of Science (21H01836, 22F22355, 22KF0058, 22K04962, 24KJ0510).
\section*{Disclosures}
Morishita, Mukherjee, El-Sadek, Makita, Yasuno: Nidek(F), Sky Technology(F), Panasonic(F), Nikon(F), Santec(F), Kao Corp.(F), Topcon(F).
Seesan, Mori, Furukawa, Matsusaka, Fukuda, Lukmanto: None.
\section*{Data availability}
The data that support the findings of this study are available from the corresponding author upon reasonable request.
The new DOCT algorithms and the DOCT simulation framework are available as open-source Python library at Refs.\@ \cite{VlivGitHub} and \cite{DoctSimuGitHub}, respectively.
\section*{Supplemental document}
See Supplement 1 for supporting content.
\bibliography{aLIVSwiftness}

\begin{thebibliography}{10}
\newcommand{\enquote}[1]{``#1''}

\bibitem{Huang1991Science}
D.~Huang, E.~A. Swanson, C.~P. Lin, J.~S. Schuman, W.~G. Stinson, W.~Chang,
  M.~R. Hee, T.~Flotte, K.~Gregory, C.~A. Puliafito, and J.~G. Fujimoto,
  \enquote{Optical coherence tomography,} {\protect\JournalTitle{Science}}
  \textbf{254}, 1178--1181 (1991).

\bibitem{Izatt1994OL}
J.~A. Izatt, E.~A. Swanson, J.~G. Fujimoto, M.~R. Hee, and G.~M. Owen,
  \enquote{Optical coherence microscopy in scattering media,}
  {\protect\JournalTitle{Opt. Lett.}} \textbf{19}, 590 (1994).

\bibitem{Beaurepaire1998OL}
E.~Beaurepaire, A.~C. Boccara, M.~Lebec, L.~Blanchot, and H.~Saint-Jalmes,
  \enquote{Full-field optical coherence microscopy,}
  {\protect\JournalTitle{Opt. Lett.}} \textbf{23}, 244 (1998).

\bibitem{Drexler1999OL}
W.~Drexler, U.~Morgner, F.~X. Kärtner, C.~Pitris, S.~A. Boppart, X.~D. Li,
  E.~P. Ippen, and J.~G. Fujimoto, \enquote{{\textit In vivo}
  ultrahigh-resolution optical coherence tomography,}
  {\protect\JournalTitle{Opt. Lett.}} \textbf{24}, 1221 (1999).

\bibitem{Dubois2004AO}
A.~Dubois, K.~Grieve, G.~Moneron, R.~Lecaque, L.~Vabre, and C.~Boccara,
  \enquote{Ultrahigh-resolution full-field optical coherence tomography,}
  {\protect\JournalTitle{Appl. Opt.}} \textbf{43}, 2874 (2004).

\bibitem{Aguirre2015Text}
A.~D. Aguirre, C.~Zhou, H.-C. Lee, O.~O. Ahsen, and J.~G. Fujimoto,
  \emph{Optical coherence microscopy} (Springer International Publishing, Cham,
  2015), chap.~28, pp. 865--911.

\bibitem{Huang2019JVE}
Y.~Huang, J.~Zou, M.~Badar, J.~Liu, W.~Shi, S.~Wang, Q.~Guo, X.~Wang,
  S.~Kessel, L.~L.-Y. Chan, P.~Li, Y.~Liu, J.~Qiu, and C.~Zhou,
  \enquote{Longitudinal morphological and physiological monitoring of
  three-dimensional tumor spheroids using optical coherence tomography,}
  {\protect\JournalTitle{J. Visualized Exp.}}  (2019).

\bibitem{SELin2021BOE}
S.-E. Lin, D.-Y. Jheng, K.-Y. Hsu, Y.-R. Liu, W.-H. Huang, H.-C. Lee, and C.-C.
  Tsai, \enquote{Rapid pseudo-{H\&E} imaging using a fluorescence-inbuilt
  optical coherence microscopic imaging system,} {\protect\JournalTitle{Biomed.
  Opt. Express}} \textbf{12}, 5139 (2021).

\bibitem{Ming2022Biosens}
Y.~Ming, S.~Hao, F.~Wang, Y.~R. Lewis-Israeli, B.~D. Volmert, Z.~Xu,
  A.~Goestenkors, A.~Aguirre, and C.~Zhou, \enquote{Longitudinal morphological
  and functional characterization of human heart organoids using optical
  coherence tomography,} {\protect\JournalTitle{Biosens. Bioelectron.}}
  \textbf{207}, 114136 (2022).

\bibitem{KYChen2023BOE}
K.~Chen, W.~Song, L.~Han, and K.~Bizheva, \enquote{Powell lens-based line-field
  spectral domain optical coherence tomography system for cellular resolution
  imaging of biological tissue,} {\protect\JournalTitle{Biomed. Opt. Express}}
  \textbf{14}, 2003 (2023).

\bibitem{KYChen2024SciRep}
K.~Chen, N.~Abbasi, A.~Wong, and K.~Bizheva, \enquote{In vivo, contactless,
  cellular resolution imaging of the human cornea with {Powell} lens based line
  field {OCT},} {\protect\JournalTitle{Sci. Rep.}} \textbf{14} (2024).

\bibitem{Apelian2016BOE}
C.~Apelian, F.~Harms, O.~Thouvenin, and A.~C. Boccara, \enquote{Dynamic full
  field optical coherence tomography: subcellular metabolic contrast revealed
  in tissues by interferometric signals temporal analysis,}
  {\protect\JournalTitle{Biomed. Opt. Express}} \textbf{7}, 1511 (2016).

\bibitem{Ren2024CB}
C.~Ren, S.~Hao, F.~Wang, A.~Matt, M.~M. Amaral, D.~Yang, L.~Wang, and C.~Zhou,
  \enquote{Dynamic contrast optical coherence tomography ({DyC-OCT}) for
  label-free live cell imaging,} {\protect\JournalTitle{Commun. Biol.}}
  \textbf{7} (2024).

\bibitem{Oldenburg2012BOE}
A.~L. Oldenburg, R.~K. Chhetri, D.~B. Hill, and B.~Button, \enquote{Monitoring
  airway mucus flow and ciliary activity with optical coherence tomography,}
  {\protect\JournalTitle{Biomed. Opt. Express}} \textbf{3}, 1978 (2012).

\bibitem{Oldenburg2015Optica}
A.~L. Oldenburg, X.~Yu, T.~Gilliss, O.~Alabi, R.~M. Taylor, and M.~A. Troester,
  \enquote{Inverse-power-law behavior of cellular motility reveals
  stromal–epithelial cell interactions in {3D} co-culture by {OCT}
  fluctuation spectroscopy,} {\protect\JournalTitle{Optica}} \textbf{2}, 877
  (2015).

\bibitem{Thouvenin2017JBO}
O.~Thouvenin, M.~Fink, and A.~C. Boccara, \enquote{Dynamic multimodal
  full-field optical coherence tomography and fluorescence structured
  illumination microscopy,} {\protect\JournalTitle{J. Biomed. Opt.}}
  \textbf{22}, 1 (2017).

\bibitem{El-Sadek2020BOE}
I.~Abd El-Sadek, A.~Miyazawa, L.~Tzu-Wei~Shen, S.~Makita, S.~Fukuda,
  T.~Yamashita, Y.~Oka, P.~Mukherjee, S.~Matsusaka, T.~Oshika, H.~Kano, and
  Y.~Yasuno, \enquote{Optical coherence tomography-based tissue dynamics
  imaging for longitudinal and drug response evaluation of tumor spheroids,}
  {\protect\JournalTitle{Biomed. Opt. Express}} \textbf{11}, 6231 (2020).

\bibitem{Park2021BOE}
S.~Park, T.~Nguyen, E.~Benoit, D.~L. Sackett, M.~Garmendia-Cedillos,
  R.~Pursley, C.~Boccara, and A.~Gandjbakhche, \enquote{Quantitative evaluation
  of the dynamic activity of {HeLa} cells in different viability states using
  dynamic full-field optical coherence microscopy,}
  {\protect\JournalTitle{Biomedical Optics Express}} \textbf{12}, 6431 (2021).

\bibitem{Scholler2019OpEx}
J.~Scholler, \enquote{Motion artifact removal and signal enhancement to achieve
  in vivo dynamic full field {OCT},} {\protect\JournalTitle{Opt. Express}}
  \textbf{27}, 19562 (2019).

\bibitem{KYFei2024BOE}
K.~Fei, Z.~Luo, Y.~Chen, Y.~Huang, S.~Li, V.~Mazlin, A.~C. Boccara, J.~Yuan,
  and P.~Xiao, \enquote{Cellular structural and functional imaging of donor and
  pathological corneas with label-free dual-mode full-field optical coherence
  tomography,} {\protect\JournalTitle{Biomed. Opt. Express}} \textbf{15}, 3869
  (2024).

\bibitem{Ling2017LSM}
Y.~Ling, X.~Yao, U.~A. Gamm, E.~Arteaga‐Solis, C.~W. Emala, M.~A. Choma, and
  C.~P. Hendon, \enquote{Ex vivo visualization of human ciliated epithelium and
  quantitative analysis of induced flow dynamics by using optical coherence
  tomography,} {\protect\JournalTitle{Lasers Surg. Med.}} \textbf{49}, 270--279
  (2017).

\bibitem{Thouvenin2017IOVS}
O.~Thouvenin, C.~Boccara, M.~Fink, J.~Sahel, M.~Pâques, and K.~Grieve,
  \enquote{Cell motility as contrast agent in retinal explant imaging with
  full-field optical coherence tomography,} {\protect\JournalTitle{Invest.
  Opthalmol. Vis. Sci.}} \textbf{58}, 4605 (2017).

\bibitem{Munter2020OL}
M.~M{\"u}nter, M.~vom Endt, M.~Pieper, M.~Casper, M.~Ahrens, T.~Kohlfaerber,
  R.~Rahmanzadeh, P.~König, G.~Hüttmann, and H.~Schulz-Hildebrandt,
  \enquote{Dynamic contrast in scanning microscopic {OCT},}
  {\protect\JournalTitle{Opt. Lett.}} \textbf{45}, 4766 (2020).

\bibitem{Leung2020BOE}
H.~M. Leung, M.~L. Wang, H.~Osman, E.~Abouei, C.~MacAulay, M.~Follen, J.~A.
  Gardecki, and G.~J. Tearney, \enquote{Imaging intracellular motion with
  dynamic micro-optical coherence tomography,} {\protect\JournalTitle{Biomed.
  Opt. Express}} \textbf{11}, 2768 (2020).

\bibitem{Xia2023Optica}
T.~Xia, K.~Umezu, D.~M. Scully, S.~Wang, and I.~V. Larina, \enquote{In vivo
  volumetric depth-resolved imaging of cilia metachronal waves using dynamic
  optical coherence tomography,} {\protect\JournalTitle{Optica}} \textbf{10},
  1439 (2023).

\bibitem{KYChen2024BOE}
K.~Chen, S.~Swanson, and K.~Bizheva, \enquote{Line-field dynamic optical
  coherence tomography platform for volumetric assessment of biological
  tissues,} {\protect\JournalTitle{Biomed. Opt. Express}} \textbf{15}, 4162
  (2024).

\bibitem{Yin2025NPJ}
Z.~Yin, B.~He, Y.~Ying, S.~Zhang, P.~Yang, Z.~Chen, Z.~Hu, Y.~Shi, R.~Xue,
  C.~Wang, S.~Wang, G.~Wang, and P.~Xue, \enquote{Fast and label-free {3D}
  virtual {H\&E} histology via active phase modulation-assisted dynamic
  full-field {OCT},} {\protect\JournalTitle{npj Imaging}} \textbf{3} (2025).

\bibitem{Hildebrandt2025OL}
H.~Schulz-Hildebrandt, J.~A. Gardecki, T.~Miller, M.~Avila-Wallace,
  E.~Villareyna-Lopez, and G.~Tearney, \enquote{Phase-sensitive dynamic
  micro-optical coherence tomography for high-speed intracellular motion
  imaging,} {\protect\JournalTitle{Optics Letters}} \textbf{50}, 4734 (2025).

\bibitem{El-Sadek2021BOE}
I.~A. El-Sadek, A.~Miyazawa, L.~T.-W. Shen, S.~Makita, P.~Mukherjee,
  A.~Lichtenegger, S.~Matsusaka, and Y.~Yasuno, \enquote{Three-dimensional
  dynamics optical coherence tomography for tumor spheroid evaluation,}
  {\protect\JournalTitle{Biomed. Opt. Express}} \textbf{12}, 6844 (2021).

\bibitem{El-Sadek2023SciRep}
I.~Abd El-Sadek, L.~T.-W. Shen, T.~Mori, S.~Makita, P.~Mukherjee,
  A.~Lichtenegger, S.~Matsusaka, and Y.~Yasuno, \enquote{Label-free drug
  response evaluation of human derived tumor spheroids using three-dimensional
  dynamic optical coherence tomography,} {\protect\JournalTitle{Sci. Rep.}}
  \textbf{13} (2023).

\bibitem{El-Sadek2024SciRep}
I.~Abd El-Sadek, R.~Morishita, T.~Mori, S.~Makita, P.~Mukherjee, S.~Matsusaka,
  and Y.~Yasuno, \enquote{Label-free visualization and quantification of the
  drug-type-dependent response of tumor spheroids by dynamic optical coherence
  tomography,} {\protect\JournalTitle{Sci. Rep.}} \textbf{14} (2024).

\bibitem{Morishita2023BOE}
R.~Morishita, T.~Suzuki, P.~Mukherjee, I.~Abd El-Sadek, Y.~Lim,
  A.~Lichtenegger, S.~Makita, K.~Tomita, Y.~Yamamoto, T.~Nagamoto, and
  Y.~Yasuno, \enquote{Label-free intratissue activity imaging of alveolar
  organoids with dynamic optical coherence tomography,}
  {\protect\JournalTitle{Biomed. Opt. Express}} \textbf{14}, 2333 (2023).

\bibitem{Mukherjee2021SciRep}
P.~Mukherjee, A.~Miyazawa, S.~Fukuda, T.~Yamashita, D.~Lukmanto, K.~Okada,
  I.~A. El-Sadek, L.~Zhu, S.~Makita, T.~Oshika, and Y.~Yasuno,
  \enquote{Label-free functional and structural imaging of liver microvascular
  complex in mice by {Jones} matrix optical coherence tomography,}
  {\protect\JournalTitle{Sci. Rep.}} \textbf{11} (2021).

\bibitem{Mukherjee2022BOE}
P.~Mukherjee, S.~Fukuda, D.~Lukmanto, T.~Yamashita, K.~Okada, S.~Makita, I.~Abd
  El-Sadek, A.~Miyazawa, L.~Zhu, R.~Morishita, A.~Lichtenegger, T.~Oshika, and
  Y.~Yasuno, \enquote{Label-free metabolic imaging of
  non-alcoholic-fatty-liver-disease ({NAFLD}) liver by volumetric dynamic
  optical coherence tomography,} {\protect\JournalTitle{Biomed. Opt. Express}}
  \textbf{13}, 4071 (2022).

\bibitem{Mukherjee2023SciRep}
P.~Mukherjee, S.~Fukuda, D.~Lukmanto, T.~H. Tran, K.~Okada, S.~Makita, I.~A.
  El-Sadek, Y.~Lim, and Y.~Yasuno, \enquote{Renal tubular function and
  morphology revealed in kidney without labeling using three-dimensional
  dynamic optical coherence tomography,} {\protect\JournalTitle{Sci. Rep.}}
  \textbf{13} (2023).

\bibitem{Guo2025arXiv}
Y.~Guo, R.~Morishita, I.~A. El-Sadek, K.~Yamazaki, S.~Sakai, P.~Mukherjee,
  Y.~Lim, C.~Bao, K.~Sugata, S.~Kasamatsu, H.~Yoshida, S.~Makita, and
  Y.~Yasuno\, \enquote{In vivo dynamic optical coherence tomography of human
  skin with hardware- and software-based motion correction,}
  {\protect\JournalTitle{arXiv:2503.21384}}  (2025).

\bibitem{Tomita2023BOE}
K.~Tomita, S.~Makita, N.~Fukutake, R.~Morishita, I.~Abd El-Sadek, P.~Mukherjee,
  A.~Lichtenegger, J.~Tamaoki, L.~Bian, M.~Kobayashi, T.~Mori, S.~Matsusaka,
  and Y.~Yasuno, \enquote{Theoretical model for en face optical coherence
  tomography imaging and its application to volumetric differential contrast
  imaging,} {\protect\JournalTitle{Biomed. Opt. Express}} \textbf{14}, 3100
  (2023).

\bibitem{VlivGitHub}
{Computational optics group at the University of Tsukuba},
  \enquote{Dynamic-optical-coherence-tomography contrast-generation library by
  computational optics group,} Open source GitHub repository (2025).
  {\url{https://github.com/ComputationalOpticsGroup/COG-dynamic-OCT-contrast-generation-library}}.

\bibitem{Przybylski2017SciRep}
A.~Przybylski, B.~Thiel, J.~Keller-Findeisen, B.~Stock, and M.~Bates,
  \enquote{Gpufit: An open-source toolkit for {GPU}-accelerated curve fitting,}
  {\protect\JournalTitle{Sci. Rep.}} \textbf{7} (2017).

\bibitem{DoctSimuGitHub}
{Computational optics group at the University of Tsukuba},
  \enquote{Dynamic-optical-coherence-tomography simulation library by
  computational optics group,} Open source GitHub repository (2025).
  {\url{https://github.com/ComputationalOpticsGroup/COG-dynamic-OCT-contrast-simulation-library}}.

\bibitem{Feng2025BOE}
Y.~Feng, S.~Fujimura, Y.~Lim, T.~Seesan, R.~Morishita, I.~Abd El-Sadek,
  P.~Mukherjee, S.~Makita, and Y.~Yasuno, \enquote{Dynamic-{OCT} simulation
  framework based on mathematical models of intratissue dynamics, image
  formation, and measurement noise,} {\protect\JournalTitle{Biomedical Optics
  Express}} \textbf{16}, 2875 (2025).

\bibitem{Spaide2015Retina}
R.~F. Spaide, J.~G. Fujimoto, and N.~K. Waheed, \enquote{Image artifacts in
  optical coherence tomography angiography,} {\protect\JournalTitle{Retina}}
  \textbf{35}, 2163--2180 (2015).

\bibitem{Nolte2024RPP}
D.~D. Nolte, \enquote{Coherent light scattering from cellular dynamics in
  living tissues,} {\protect\JournalTitle{Reports on Progress in Physics}}
  \textbf{87}, 036601 (2024).

\bibitem{Seesan2021BOE}
T.~Seesan, I.~Abd El-Sadek, P.~Mukherjee, L.~Zhu, K.~Oikawa, A.~Miyazawa,
  L.~T.-W. Shen, S.~Matsusaka, P.~Buranasiri, S.~Makita, and Y.~Yasuno,
  \enquote{Deep convolutional neural network-based scatterer density and
  resolution estimators in optical coherence tomography,}
  {\protect\JournalTitle{Biomed. Opt. Express}} \textbf{13}, 168 (2021).

\bibitem{Seesan2024BOE}
T.~Seesan, P.~Mukherjee, I.~A. El-Sadek, Y.~Lim, L.~Zhu, S.~Makita, and
  Y.~Yasuno, \enquote{Optical-coherence-tomography-based deep-learning
  scatterer-density estimator using physically accurate noise model,}
  {\protect\JournalTitle{Biomed. Opt. Express}} \textbf{15}, 2832--2848 (2024).

\bibitem{Feng2025BiOS}
F.~Yuanke, F.~Shumpei, L.~Yiheng, S.~Thitiya, M.~Rion, E.-S. Ibrahim, Abd,
  M.~Pradipta, and Y.~Yoshiaki, \enquote{Dynamic {OCT} simulator based on
  mathematical models of intratissue dynamics, image formation, and measurement
  noise,} in \emph{Optical Coherence Tomography and Coherence DomainOptical
  Methods in Biomedicine XXIX,}  (Proc. SPIE, 2025).

\bibitem{Fujimura2025BiOS}
F.~Shumpei, E.-S. Ibrahim, Abd, M.~Rion, F.~Yuanke, and Y.~Yoshiaki,
  \enquote{Intracellular activity-type and parameter estimation from dynamic
  opticalcoherence tomography signals,} in \emph{Optical Coherence Tomography
  and Coherence DomainOptical Methods in Biomedicine XXIX,}  (Proc. SPIE,
  2025).

\bibitem{Li2017BOE}
E.~Li, S.~Makita, Y.-J. Hong, D.~Kasaragod, and Y.~Yasuno,
  \enquote{Three-dimensional multi-contrast imaging of {\textit{in vivo}} human
  skin by {Jones} matrix optical coherence tomography,}
  {\protect\JournalTitle{Biomed. Opt. Express}} \textbf{8}, 1290 (2017).

\bibitem{Miyazawa2019BOE}
A.~Miyazawa, S.~Makita, E.~Li, K.~Yamazaki, M.~Kobayashi, S.~Sakai, and
  Y.~Yasuno, \enquote{Polarization-sensitive optical coherence elastography,}
  {\protect\JournalTitle{Biomed. Opt. Express}} \textbf{10}, 5162 (2019).

\bibitem{Costa2016BA}
E.~C. Costa, A.~F. Moreira, D.~de~Melo-Diogo, V.~M. Gaspar, M.~P. Carvalho, and
  I.~J. Correia, \enquote{{3D} tumor spheroids: an overview on the tools and
  techniques used for their analysis,} {\protect\JournalTitle{Biotechnol.
  Adv.}} \textbf{34}, 1427--1441 (2016).

\bibitem{SchulzHildebrandt2024FMN}
H.~Schulz-Hildebrandt, S.~Spasic, F.~Hou, K.-C. Ting, S.~Batts, G.~Tearney, and
  K.~M. Stankovic, \enquote{Dynamic micro-optical coherence tomography enables
  structural and metabolic imaging of the mammalian cochlea,}
  {\protect\JournalTitle{Frontiers in Molecular Neuroscience}} \textbf{17}
  (2024).

\bibitem{Zhou2021AOP}
K.~C. Zhou, R.~Qian, A.-H. Dhalla, S.~Farsiu, and J.~A. Izatt, \enquote{Unified
  k-space theory of optical coherence tomography,}
  {\protect\JournalTitle{Advances in Optics and Photonics}} \textbf{13}, 462
  (2021).

\bibitem{Noah2025BOE}
N.~Heldt, T.~Monfort, R.~Morishita, R.~Schönherr, O.~Thouvenin, I.~El-Sadek,
  P.~König, G.~Hüttmann, K.~Grieve, and Y.~Yasuno\, \enquote{A guide to
  dynamic {OCT} data analysis,} {\protect\JournalTitle{Biomedical Optics
  Express}} \textbf{16} (2025).

\end{thebibliography}

\pagebreak
\title{Supplementary Material: Supplement 1}
\setcounter{figure}{0}
\renewcommand\thefigure{S\arabic{figure}}   
\setcounter{table}{0}
\renewcommand\thetable{S\arabic{table}}   
\setcounter{section}{0}
\renewcommand\thesection{S\arabic{section}}  

\section{Analytic field of numerical simulation}
\begin{figure}[htbp]
	\centering
	\fbox{\includegraphics{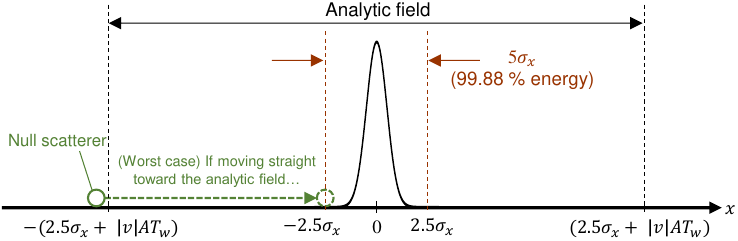}}
	\caption{Schematic depiction of the analytic field size and complex PSF for $x$ direction. }
	\label{fig:fieldSize}
\end{figure}

Figure \ref{fig:fieldSize} schematically show the analytic field size for $x$ direction.

\section{Terminology}
The terminology and abbreviations used in this paper are summarized in Table \ref{tab:terminology}.
\begin{table}
	\caption{Terminology and abbreviations.}
	\label{tab:terminology}
	\centering
	\small
	\begin{tabular}{cp{8cm}}
		\hline
		\centering
		Abbreviation & Meaning \\
		\hline
		DOCT & Dynamic optical coherence tomography \\
		$\Atw$ & Acquisition time window of full data \\
		$\tw$ & Time window of data subset to compute LIV \\
		$\VAtw$ & Acquisition time window for virtual frame-decreased data\\
		LIV & Logarithmic intensity variance\\
		OCDS & OCT correlation decay speed\\
		aLIV & Authentic LIV\\
		LIV curve & Sequential LIV over $\tw$ at a single location\\
		$a$ & Saturation level; first fitting parameter\\
		$\tau$ & Time constant; second fitting parameter\\
		PSF & Point spread function\\
		FOV & Field of view\\
		PTX & Paclitaxel; one type of anti-cancer drugs\\
		\DSR & Dynamic-scatterer ratio\\
		ROI & Region of interest\\
		\hline
	\end{tabular}
\end{table}

\end{document}